\documentclass[useAMS,usenatbib]{mn2e}
\usepackage{subfig}
\usepackage{graphicx}

\title[An infrared view of AGN feedback]{An infrared view of AGN feedback in a type-2 quasar: the case of the Teacup galaxy}
\author[C. Ramos Almeida et al.]
{\parbox{\textwidth}{C. Ramos Almeida$^{1,2}$\thanks{Ram\'on y Cajal fellow. E-mail: cra@iac.es},
J. Piqueras L\' opez$^{3,4,5}$,
M. Villar-Mart\' in$^{3,4}$, 
P. S. Bessiere$^{6}$
}\vspace{0.4cm}\\
\parbox{\textwidth}{$^{1}$Instituto de Astrof\' isica de Canarias, Calle V\' ia L\'actea, s/n, E-38205, La Laguna, Tenerife, Spain\\
$^{2}$Departamento de Astrof\' isica, Universidad de La Laguna, E-38206, La Laguna, Tenerife, Spain\\
$^{3}$Centro de Astrobiolog\' ia (INTA-CSIC), Carretera de Ajalvir, km 4, E-28850, Torrej\'on de Ardoz, Madrid, Spain\\
$^{4}$Astro-UAM, UAM, Unidad Asociada CSIC, Facultad de Ciencias, Campus de Cantoblanco, E-28049, Madrid, Spain\\
$^{5}$Department of Astrophysics, University of Oxford, Keble Road, Oxford OX1 3RH, UK\\ 
$^{6}$Instituto de Astrof\' isica, Facultad de F\' isica, Pontificia Universidad Cat\'olica de Chile, Casilla 306, Santiago 22, Chile\\ 
}}

\begin{document}

\date{}

\pagerange{\pageref{firstpage}--\pageref{lastpage}} \pubyear{2017}

\maketitle

\label{firstpage}

\begin{abstract}
We present near-infrared integral field spectroscopy data obtained with VLT/SINFONI 
of ``the Teacup galaxy''.
The nuclear K-band (1.95--2.45~\micron) spectrum of this radio-quiet type-2 quasar reveals a blueshifted broad component of 
FWHM$\sim$1600-1800 km~s$^{-1}$ in the hydrogen recombination lines 
(Pa$\alpha$, Br$\delta$, and Br$\gamma$) and also in the coronal line [Si VI]$\lambda$1.963 \micron. Thus the data
confirm the presence of the nuclear ionized outflow previously detected in the optical and reveal its 
coronal counterpart. 
Both the ionized and coronal nuclear outflows are resolved, with seeing-deconvolved full widths at half maximum of 
1.1$\pm$0.1 and 0.9$\pm$0.1 kpc along PA$\sim$72\degr--74\degr. This orientation is almost coincident with the radio axis (PA=77\degr),
suggesting that the radio jet could have triggered the nuclear outflow. 
In the case of the H$_2$ lines we do not require a broad component to reproduce the profiles, but the narrow lines 
are blueshifted by $\sim$50 km~s$^{-1}$ on average from the galaxy systemic velocity. This could be an indication of the 
presence of a nuclear molecular outflow, although the bulk of the H$_2$ emission in the inner $\sim$2\arcsec~($\sim$3 kpc) of the galaxy 
follows a rotation pattern. 
We find evidence for kinematically disrupted gas (FWHM$>$250 km~s$^{-1}$) at up to 5.6 kpc from the AGN, which can be naturally explained 
by the action of the outflow.
The narrow component of [Si VI] is redshifted with respect to the 
systemic velocity, unlike any other emission line in the K-band spectrum. This 
indicates that the region where the coronal lines are produced is not co-spatial with the narrow line region.
\end{abstract}





\section{Introduction}
\label{intro}


Cosmological simulations require active galactic nuclei (AGN) feedback to regulate black hole and galaxy 
growth \citep{diMatteo05,Croton06}. This process occurs when the intense radiation produced by the active nucleus sweeps out
and/or heats the interstellar gas, quenching star formation and therefore producing a more realistic number of massive galaxies in
the simulations (see \citealt{Fabian12} for a review). 
Two major modes of AGN feedback are identified. Radio- or kinetic-mode feedback dominates in galaxy clusters and groups, where
jet-driven radio bubbles heat the intra-cluster medium. This type of feedback is generally associated with powerful radio galaxies. 
Quasar- or radiative-mode feedback consists of AGN-driven winds of ionized, neutral, and molecular gas \citep{Fabian12,Fiore17}. 
However, such a clear distinction between the two modes of feedback can be somewhat misleading. This is because it has been shown, 
through the detection of nuclear outflows, that radiative-mode feedback also acts in radio-galaxies (see e.g. \citealt{Emonts16} and 
references therein), whilst the presence of jets in quasars that are deemed to be radio-quiet can lead to faster and more turbulent 
AGN-driven winds \citep{Mullaney13,Zakamska14}. Therefore, it is over-simplistic to consider the impact each mode of AGN-feedback 
has on its host galaxy in isolation.  


One of the most efficient ways to identify the imprint of outflows in large AGN samples at any
redshift is to search for them in the warm ionized phase in the optical range (e.g. [O III]$\lambda$5007 \AA).
Indeed, during recent years it has become clear that ionized outflows are a ubiquitous phenomenon in type-2 quasars 
(QSO2s) at z$\la$0.7 \citep{Villar11,Villar14,Liu13,Harrison14,Karouzos16}. QSO2s are excellent laboratories to search for
outflows and study their impact on their host galaxies, as the AGN continuum and the broad components of 
the permitted lines produced in the broad-line region (BLR) are obscured. Fast motions are often measured in QSO2s, 
with full-widths at half maximum (FWHM) $>$1000 km~s$^{-1}$ and typical velocity shifts (V$_s$) of hundreds km~s$^{-1}$. 
These outflows are likely triggered by AGN-related
processes and they originate in the high-density regions (n$_e\ge 10^3~cm^{-3}$) within the central kiloparsecs of 
the galaxies. Optical integral field spectroscopy (IFS) studies have shown that these outflows can extend up to 
$\sim$15 kpc from the AGN \citep{Humphrey10,Liu13,Harrison14}. However, these results have been recently questioned, 
as the reported outflow extents could be overestimated due to seeing smearing effects 
\citep{Karouzos16,Villar16,Husemann16}.



Now that ionized outflows have been identified as a common process in QSO2,
the next goal is to investigate their impact on other gaseous phases, such as the molecular and coronal phases. Since H$_2$ 
is the fuel required to form stars and feed the SMBH, the impact of the outflows in this gaseous phase is what might truly
affect how systems evolve. Detecting coronal outflows is also interesting because, due to the 
high ionization potentials (IP$\ga$100 eV; \citealt{Mullaney09,Rodriguez11,Landt15}) of these lines, they are unequivocally
associated with nuclear activity. Coronal lines have intermediate widths between those of the broad and the narrow
emission lines (FWHM$\sim$500--1500 km~s$^{-1}$) and are generally blueshifted and/or more asymmetric than 
lower-ionization lines \citep{Penston84,Appenzeller91,Rodriguez11}. This indicates that either coronal lines are produced 
in an intermediate region between the narrow-line region (NLR) and the BLR \citep{Brotherton94,Mullaney08,Denney12}, and/or 
are related to 
outflows \citep{Muller06,Muller11}.

The near-infrared (NIR) range, and particularly the K-band, allows us to trace 
outflow signatures in the molecular, ionized and coronal phases simultaneously. In addition, because ionized outflows
in QSO2 are heavily reddened \citep{Villar14}, observing them in the NIR permits us to
penetrate through the dust screen and trace the regions closer to the base of the outflow.

The rest-frame NIR spectrum of QSO2s at z$<$0.7 has not been fully characterised yet. To the best of our knowledge,
this has been done for one QSO2 so far: Mrk\,477 at z=0.037 \citep{Villar15}. Additionally, a NIR spectrum of the QSO2 
SDSS J1131+1627 at z=0.173 was presented in \citet{Rose11}, but only Pa$\alpha$ was detected. 
Here we explore the NIR spectrum of the QSO2 SDSS J143029.88+133912.0 (J1430+1339; at $z$=0.0852).


\subsection{The Teacup galaxy}

\begin{figure*}
\centering
\includegraphics[width=12cm]{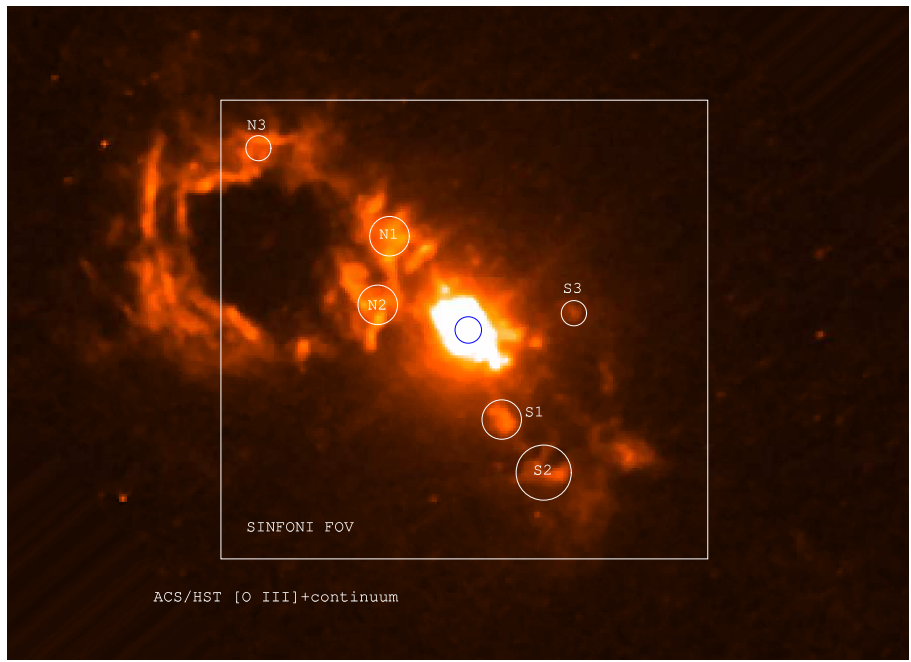}
\caption{HST/ACS [O III]$\lambda$5007 \AA~image of the central 12.4\arcsec$\times$17.2\arcsec~(19.7$\times$27.4 kpc$^2$) of the Teacup. 
Note the smaller FOV covered by SINFONI (9.2\arcsec$\times$8.7\arcsec$\simeq$ 14.6$\times$13.8 kpc$^2$). The circles
correspond to the different apertures chosen to extract the nuclear (blue central circle), eastern bubble (N1, N2, and N3), and western bubble 
NIR spectra (S1, S2, S3). North is up and east to the left.}
\label{fig1}
\end{figure*}

According to its [O III] luminosity (5$\times$10$^{42}~erg~s^{-1}$ = 10$^{9.1}L_{\sun}$; \citealt{Reyes08}), 
J1430+1339 is a luminous QSO2, and considering its position in the 1.4GHz--[O III] luminosity plane \citep{Lal10} 
it is classified as radio-quiet 
(L$_{1.4GHz}=5\times 10^{23}~W Hz^{-1}$; Harrison et al. 2015, hereafter \citealt{Harrison15}).
Nonetheless, it is a factor of 10 above the radio-FIR correlation found for star-forming galaxies \citep{Villar14,Harrison14}, 
which makes it a ``radio excess source''.
The host galaxy shows clear signatures of a past interaction with another galaxy, in the form of shells, tails and chaotic 
dust lanes \citep{Keel15}.
J1430+1339 was nicknamed ``the Teacup galaxy'' because of the peculiar appearance of its extended emission-line region (EELR) 
in SDSS and HST images \citep{Keel12,Keel15}. This EELR is dominated by a filamentary bubble to the northeast (NE) with a radial 
extent of $\sim$12 kpc measured from the nucleus (see Figure \ref{fig1}). In the opposite direction there is another
knotty emission-line structure resembling a fan extending up to $\sim$7 kpc. The Teacup has been proposed as a fading AGN candidate 
\citep{Gagne14}.

The NE emission-line bubble coincides with the radio-continuum structure detected in VLA maps \citet{Harrison15}. 
These radio maps also show another radio bubble extending $\sim$10 kpc to the west, as well as two compact 
radio structures: a brighter one coincident with the AGN position, and a fainter one 
located $\sim$0.8 kpc northeast from the AGN (PA$\sim$60\degr), identified by \citealt{Harrison15} as high-resolution B (HR-B) region. 
According to the latter authors, this HR-B structure would be co-spatial with the base of the ionized nuclear outflow first 
reported by \citet{Villar14} and \citet{Harrison14} using the SDSS spectrum and IFU spectroscopy respectively. At the position 
of HR-B the outflow has an observed velocity of -740 km~s$^{-1}$ relative to the narrow component of [O III]$\lambda$5007 \AA, 
which \citealt{Harrison15} interpreted as gas accelerated by jets or quasar winds at that location. The latter authors also 
speculated that these jets/winds would be driving the 10--12 kpc radio bubbles.
 
Here we study the nuclear and extended NIR emission of the Teacup using seeing-limited SINFONI K-band data. 
In Section 2 we describe the observations and data reduction, in Section 3 we present the results on the 
nuclear and extended emission of the galaxy, and in Section 4 we discuss the observations and their implications. 
Throughout this paper we assume 
a cosmology with H$_0$=71 km~s$^{-1}$~Mpc$^{-1}$, $\Omega_m$ = 0.27, and $\Omega_{\Lambda}$ =0.73. At the redshift of the 
galaxy (z=0.0852), the spatial scale is 1.591 kpc~arcsec$^{-1}$. 

\section{Observations and Data Reduction}
\label{observations}

We obtained K-band (1.95--2.45 \micron) observations of the Teacup with SINFONI on the 8 m Very Large Telescope (VLT). 
The data were taken during the night of 2015 March 7th in service mode (Program ID: 094.B-0189(A)) with a total on-source time
of 1800 s and at an airmass of 1.3--1.4. Due to the strong and rapid variation of the IR sky emission, 
the observations were split into short exposures of 
300 s each, following a jittering O-S-S-O pattern for sky and on-source frames.

The observing conditions were clear and the seeing variation over the on-source observing period was small according to the 
DIMM seeing monitor\footnote{http://archive.eso.org/wdb/wdb/asm/historical$_-$ambient$_-$paranal/query} (median optical seeing
FWHM=1.05\arcsec, standard deviation=0.21\arcsec, and standard error=0.02\arcsec). To calculate the seeing FWHM in the K-band 
we used the photometric standard star observed immediately after the target, which appears slightly 
elongated along PA=88.3$\pm$0.1\degr, with a maximum FWHM=0.58\arcsec. Along the minor axis, the FWHM=0.46\arcsec. 
Therefore, the seeing error is dominated by the shape of the PSF rather than by the seeing variation during the observations. 
By averaging the maximum and 
minimum values of the FWHM measured for the star and by adding the seeing variation and PSF shape errors in quadrature, we get a seeing  
FWHM=0.52$\pm$0.06\arcsec~($\sim$830 pc resolution).

We used the 0.125\arcsec$\times$0.250\arcsec~pixel$^{-1}$ 
configuration, which yields a field-of-view (FOV) of 8\arcsec$\times$8\arcsec~per single exposure. Due to the jittering process, 
the effective FOV in the case of our target is $\sim$9\arcsec$\times$9\arcsec~($\sim$14$\times$14 kpc$^2$). The spectral resolution
in the K-band is R$\sim$3300 ($\sim$75 km~s$^{-1}$) and the instrumental broadening, as measured from the OH sky lines, 
is 6.0$\pm$0.5 \AA~with a dispersion of 2.45 \AA~pixel$^{-1}$. 

For the reduction of the data, we used the ESO pipeline ESOREX (version 3.8.3) and our own IDL routines for the telluric
correction and flux calibration (see \citealt{Piqueras12}). 
We applied the usual calibration corrections of dark subtraction, flat fielding, detector linearity, 
geometrical distortion, wavelength calibration, and subtraction of the sky emission to the individual frames. 
The individual cubes from each exposures were then combined into a single data cube. 
To estimate the uncertainty in the wavelength calibration we used the atmospheric OH lines, from which we measured an error
of 7.8 km~s$^{-1}$.

The flux calibration was performed in two steps. First, to obtain the atmospheric transmission curves, we extracted the 
spectra of the standard stars with an aperture of 5$\sigma$ of the best 2D Gaussian fit of a collapsed image. The spectra 
were then normalised by a blackbody profile of the appropiate T$_{eff}$, taking the most relevant absorption spectral 
features of the stars into account. The result is a sensitivity function that accounts for the atmospheric transmission. 
Second, we flux-calibrated the spectra of the stars using their 2MASS K-band magnitudes. Every individual cube was then 
divided by the sensitivity function and multiplied by the conversion factor to obtain a fully-calibrated data cube. The 
uncertainty in the flux calibration is $\sim$15\%. We refer the reader to \citet{Piqueras12,Piqueras16} for further details 
on the data reduction.

\section{Results}

\subsection{Nuclear spectrum}

We extracted two K-band spectra of the nuclear region of the Teacup in two circular apertures of 0.5\arcsec~and 
1.25\arcsec~diameter ($\sim$0.8 and 2 kpc respectively), centred 
at the maximum of the Pa$\alpha$ emission (see Section \ref{nuclear} for details). The minimum aperture was
chosen to match the spatial resolution set by the seeing (FWHM=0.52$\pm$0.06\arcsec).
In the following we will refer to the spectrum extracted in this aperture as the nuclear spectrum. 

\subsubsection{Continuum shape}
\label{nuclear}

\begin{figure*}
\centering
{\par\includegraphics[width=6.1cm,angle=90]{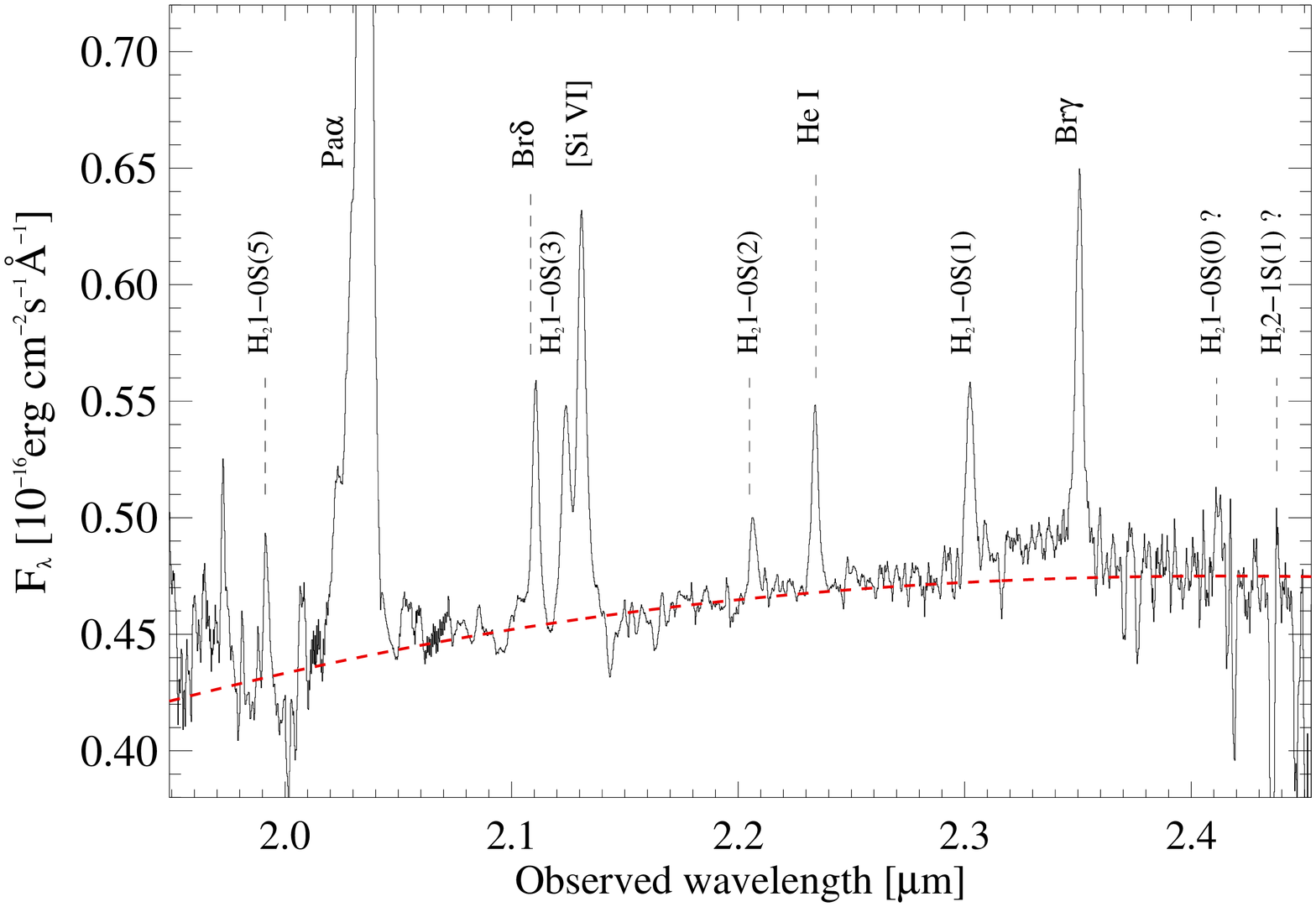}
\includegraphics[width=6.1cm,angle=90]{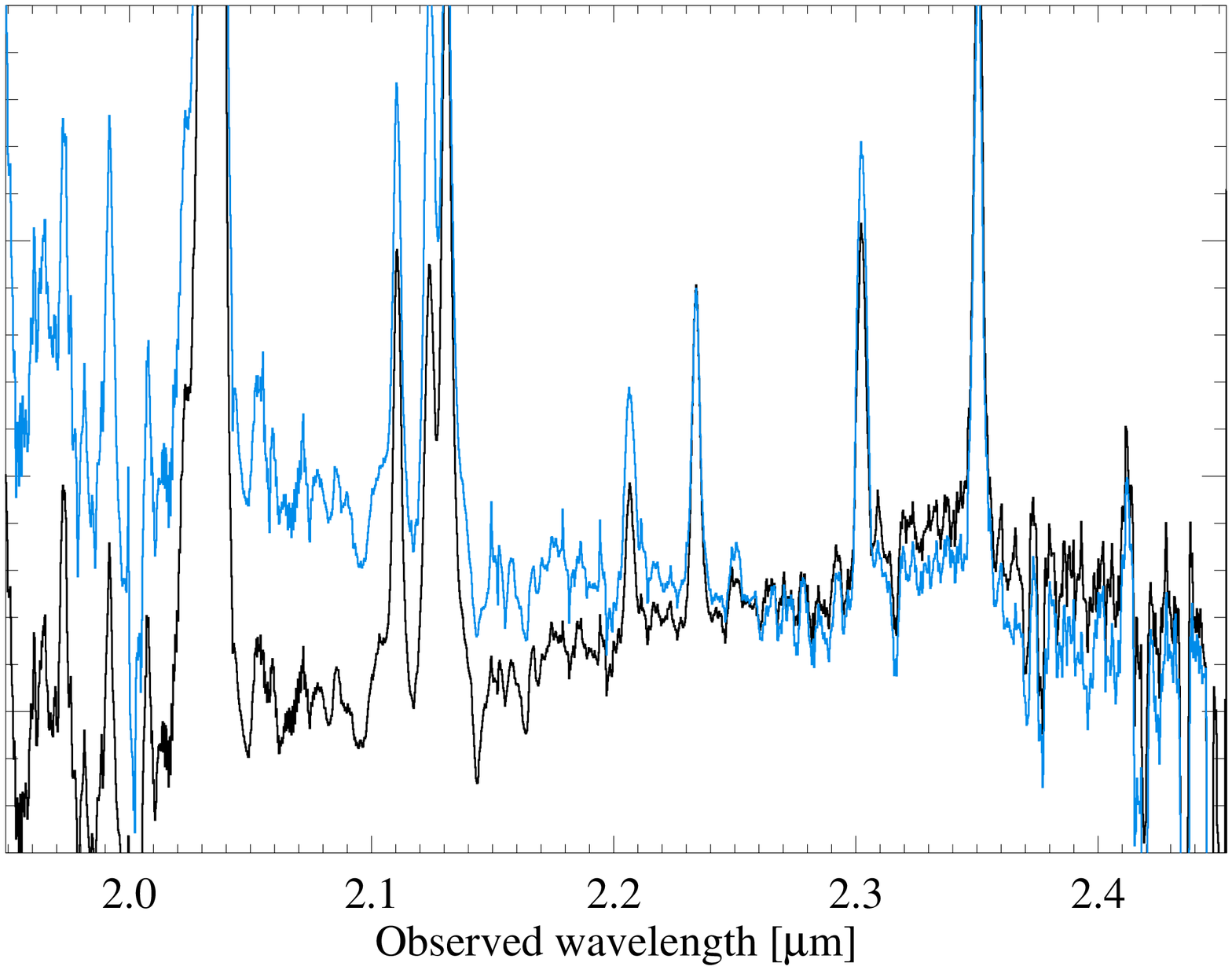}\par}
\caption{Left: Flux-calibrated nuclear spectrum of the Teacup, extracted in a circular aperture of 0.5\arcsec~diameter and 
smoothed using a 6 pixels boxcar.
The most prominent emission lines are labelled. The molecular lines H$_2$1-0S(0) and 2-1S(1) are only marginally detected. 
The red dashed line corresponds to a blackbody of T=1200 K. Right: Spectra of the
Teacup extracted in apertures of 0.5\arcsec~(black line) and 1.25\arcsec~diameter (blue line), 
normalized at 2.27 \micron. Note the rise of the continuum at shorter wavelengths with increasing aperture.}
\label{fig2}
\end{figure*}

Figure \ref{fig2} shows the observed nuclear spectrum of the Teacup with the emission lines labelled. 
We extracted the spectra centred at the peak of the ionized gas emission (as traced by Pa$\alpha$) 
because it does not coincide with the maximum of the K-band continuum emission. 
The peaks of the ionized gas and continuum emission are spatially offset by
0.125\arcsec~($\sim$200 pc; i.e. the size of one spaxel) with PA=0\degr. 
It is the case that 0.125\arcsec~is 1/4 the seeing size 
and thus we cannot resolve two spectra spatially offset by 0.125\arcsec.

The continuum slope 
rises towards the red, showing a maximum at $\sim$2.35 \micron. The red dashed line in the left panel of 
Figure \ref{fig2} corresponds to a blackbody of T=1200 K, which better reproduces the K-band nuclear continuum of the Teacup. 
This K-band spectral shape is not common in type-2 AGN, which generally show the opposite slope, 
but it has been reported for a few Seyfert 2 galaxies (e.g. NGC\,7674 -- 
\citealt{Riffel06}; Mrk\,348 -- \citealt{Ramos09}; NGC\,4472 and NGC\,7743 -- \citealt{Burtscher15}) and it has been interpreted 
as emission from AGN-heated nuclear dust near the sublimation temperature. 
In the case of the Teacup, considering the relatively large area probed by the nuclear spectrum ($\sim$830 pc diameter),
we are likely detecting hot polar dust within the ionization cones.  

The change of the continuum slope with increasing aperture, as shown in the right 
panel of Figure \ref{fig2} is also noteworthy. The red excess is only observed in the 
nuclear spectrum, it then flattens if we consider intermediate apertures and the slope becomes negative in the case 
of the large aperture spectrum (1.25\arcsec~diameter). The latter spectral shape resembles the typical K-band 
spectrum of Seyfert 2 galaxies \citep{Riffel06,Ramos09,Burtscher15}. This change of slope with increasing aperture is due to the
extra contribution from stellar light included in the larger apertures.

\subsubsection{Emission line spectrum}
\label{lines}

\begin{figure}
\centering
{\par\includegraphics[width=5.9cm,angle=90]{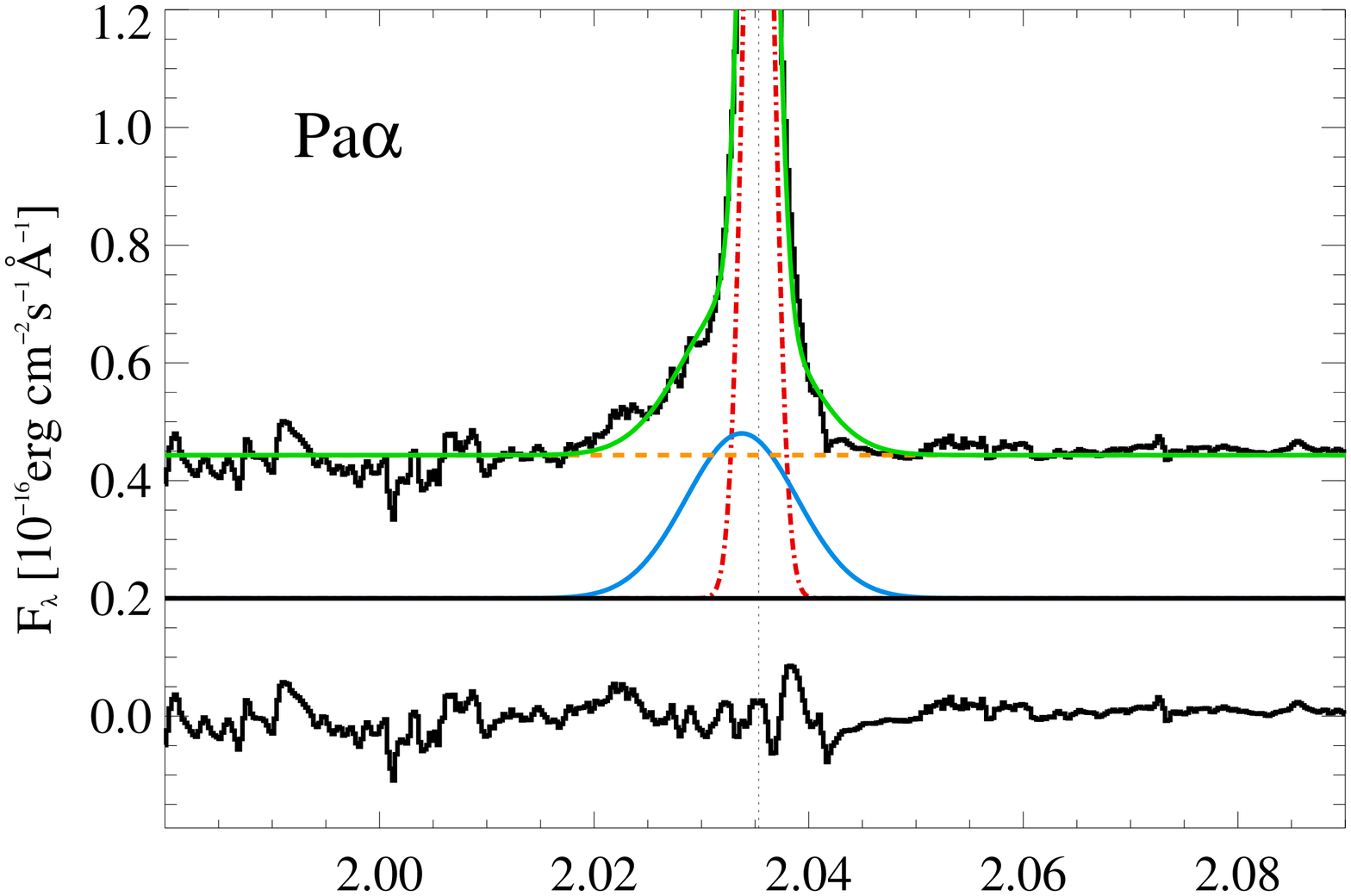}
\includegraphics[width=5.9cm,angle=90]{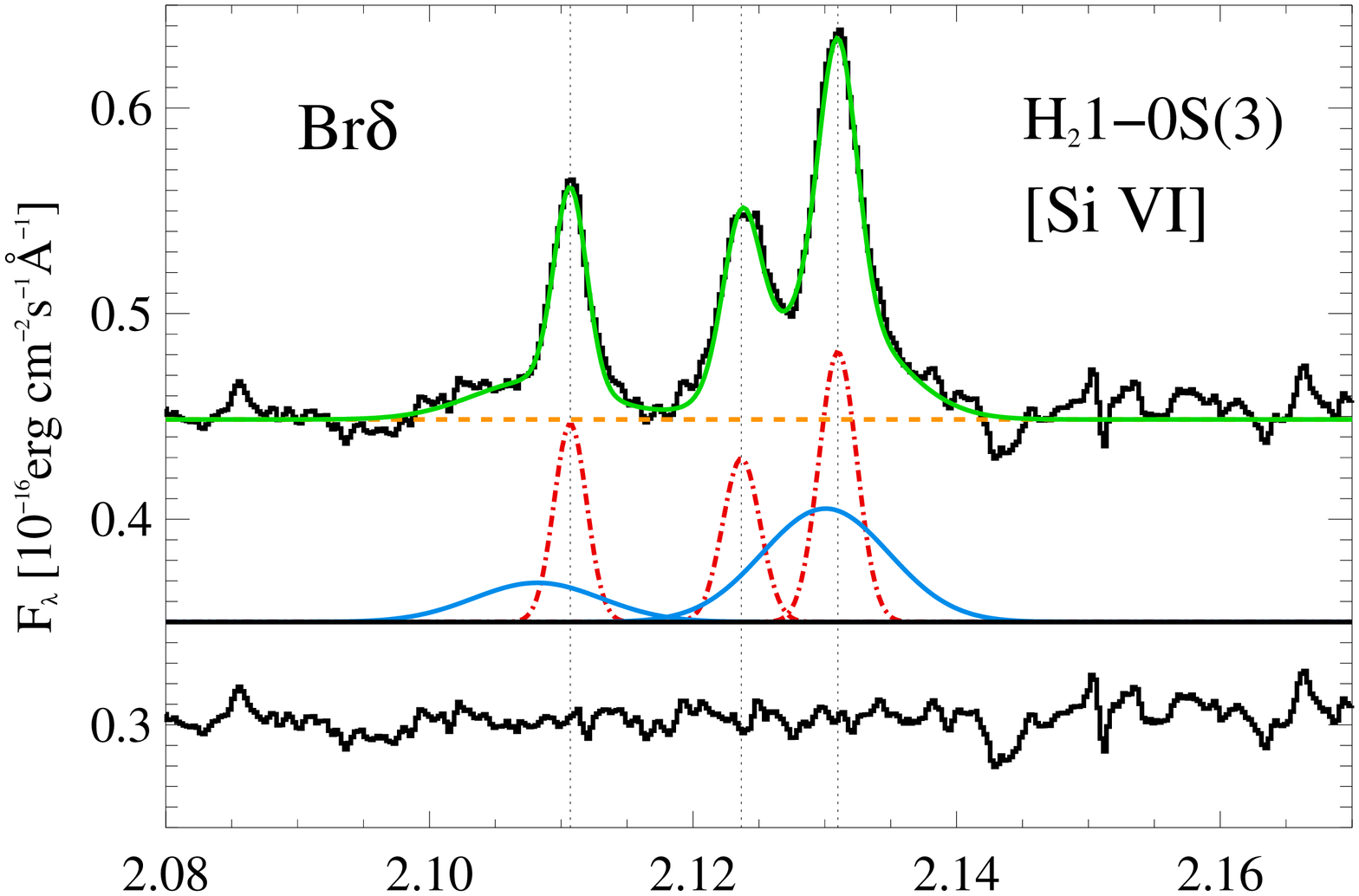}
\includegraphics[width=5.9cm,angle=90]{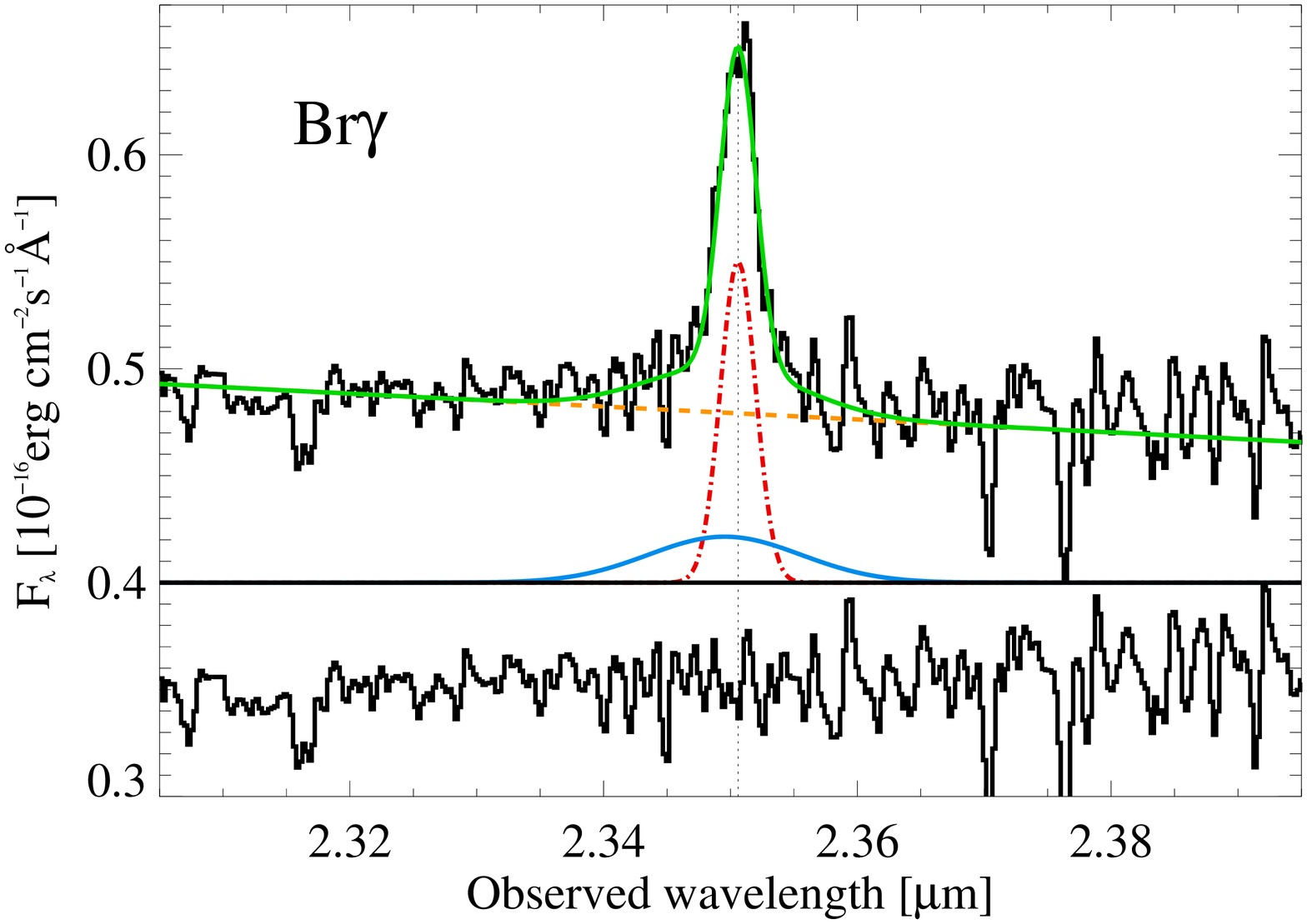}\par}
\caption{Line profiles showing a blueshifted broad component in the nuclear spectrum of the Teacup. Corresponding fits are shown as 
solid green lines. Solid blue and dot-dashed red Gaussians are the broad and narrow line components respectively,
and the orange dashed line is the continuum. The Gaussians have been vertically shifted for displaying purposes. The insets at 
the bottom of each panel are the residuals, which in the case of the middle and bottom panels, have also been scaled vertically.}
\label{fig3}
\end{figure}

By far the most prominent emission-line feature in the nuclear spectrum of the galaxy is Pa$\alpha$
(see Figure \ref{fig2}), followed by Br$\delta$, He I$\lambda$2.060, Br$\gamma$, and the coronal line [Si VI]$\lambda$1.963 
(all $\lambda$ given in \micron~unless otherwise specified). We also detect several H$_2$ emission lines, indicative of the presence of a 
nuclear molecular gas reservoir. We fitted the nuclear emission-line spectrum 
with Gaussian profiles using the Starlink program {\sc dipso}. Since the [Si VI] and H$_2$ 1-0S(3) lines are blended (see central
panel of Figure \ref{fig3}) we fixed the FWHM of H$_2$ 1-0S(3) to match those of the other H$_2$ lines to enable us to obtain a reliable
fit. In Table \ref{tab1} we report the FWHMs corrected for instrumental broadening, velocity shifts (V$_s$) and fluxes resulting 
from our fits with {\sc dipso} with their correspoding errors. The uncertainties in V$_s$ include the wavelength
calibration error (7.8 km~s$^{-1}$ as measured from the sky spectrum) and the individual fit uncertainties provided by {\sc dipso}. In the case
of the fluxes, the errors have been determined by adding quadratically the flux calibration error (15\%) and the fit uncertainties.

We require two Gaussians to reproduce the hydrogen 
recombination lines and the [Si VI]$\lambda$1.963 line profiles. This includes a narrow component of FWHM$\sim$400-460
km~s$^{-1}$ and a broad blueshifted
component of FWHM$\sim$1600--1800 km~s$^{-1}$. We identify this broad component with the nuclear outflow reported from optical
spectroscopy by \citet{Villar14} and \citealt{Harrison15}. 
We discard the possibility of a BLR origin because
the broad components are significantly blueshifted from the narrow component. Additionally not only are they detected 
in the permitted lines, 
but also in the [Si VI] coronal line. 
 
\begin{table}
\centering
\begin{tabular}{lccc}
\hline
\hline
 &\multicolumn{3}{c}{Nuclear spectrum} \\
Line & FWHM & V$_s$ & Line flux $\times 10^{15}$ \\
& (km~s$^{-1}) $  & (km~s$^{-1}$) & (ergs~cm$^{-2}$~s$^{-1}$)\\
\hline
Pa$\alpha$      	 &  434$\pm$7	&    0$\pm$8   & 5.96$\pm$0.90   \\
Pa$\alpha$ (b)        	 & 1794$\pm$93  & -234$\pm$35  & 3.64$\pm$0.56   \\
Br$\delta$ 	 	 &  400$\pm$23  &   -5$\pm$11  & 0.30$\pm$0.05   \\
Br$\delta$ (b)	 	 & 1620$\pm$164 & -361$\pm$109 & 0.23$\pm$0.05   \\
Br$\gamma$       	 &  406$\pm$59  &    2$\pm$15  & 0.52$\pm$0.12   \\
Br$\gamma$ (b)      	 & 1807$\pm$1162& -133$\pm$209 & 0.32$\pm$0.12   \\
He I                 	 &  416$\pm$24  &    1$\pm$12  & 0.27$\pm$0.04   \\
\hline
$[Si~VI]$ 		 &  463$\pm$22  &   54$\pm$11  & 0.47$\pm$0.08   \\
$[Si~VI]$ (b)		 & 1594$\pm$124 &  -77$\pm$55  & 0.67$\pm$0.11   \\
\hline
H$_{2}$ 1-0S(5)	 	 &  401$\pm$114 &  -63$\pm$37  & 0.22$\pm$0.06   \\
H$_{2}$ 1-0S(3)	 	 &  460 	&  -89$\pm$14  & 0.28$\pm$0.05   \\
H$_{2}$ 1-0S(2)		 &  452$\pm$52  &  -38$\pm$23  & 0.13$\pm$0.02   \\
H$_{2}$ 1-0S(1)		 &  477$\pm$33  &  -14$\pm$17  & 0.32$\pm$0.05   \\
H$_{2}$ 1-0S(0)		 & \dots	& \dots	       & :0.06	         \\
H$_{2}$ 2-1S(1)		 & \dots	& \dots	       & :0.06	         \\
\hline	   					 			    																																				      
\end{tabular}	
\caption{Emission lines detected in the nuclear spectrum of the Teacup 
(0.5\arcsec~diameter). FWHMs are corrected from instrumental broadening, 
and the velocity shifts (V$_s$) have been calculated relative to the central 
$\lambda$ of the narrow component of Pa$\alpha$. 
In the case of the molecular lines 1-0S(0) and 2-1S(1), the reported 
fluxes correspond to upper limits at 3$\sigma$. The FWHM of the H$_2$ 1-0S(3) line was fixed 
to obtain a reliable fit.}
\label{tab1}
\end{table}

In Figure \ref{fig3} we show the profiles and corresponding fits of the emission lines showing blueshifted broad components in the
nuclear spectrum, namely Pa$\alpha$, Br$\delta$, [Si VI], and Br$\gamma$. In the case of Pa$\alpha$ we fitted a broad
component of FWHM=1800$\pm$90 km~s$^{-1}$ with V$_s$=-234$\pm$35 km~s$^{-1}$
relative to the central wavelength of the narrow component ($\lambda_c$=20353.39$\pm$0.54 \AA, giving $z$=0.08516$\pm$0.00003). 
For Br$\delta$ and Br$\gamma$ we fitted blueshifted broad components consistent with that of Pa$\alpha$ within the 
uncertainties (see Table \ref{tab1}). 

The coronal line [Si VI] also shows a broad component of FWHM=1600$\pm$120 km~s$^{-1}$ blueshifted with respect to the 
narrow [Si VI] emission component. 
It is noteworthy that the narrow component is redshifted by V$_s$=54$\pm$11 km~s$^{-1}$ from the narrow Pa$\alpha$ line, 
which is not observed in any other emission line in the K-band spectrum of the Teacup. As explained in Section
\ref{intro},
coronal lines are generally blueshifted and slightly broader than lower ionization emission lines (e.g. \citealt{Rodriguez11}). 
In the case of the Teacup,
we detect the blueshifted broad component associated with the outflow, and a redshifted narrow component whose FWHM
is the same as those of the recombination lines. 
This redshifted narrow component
suggests that the coronal region and the region where the narrow core of the hydrogen lines is produced are different. 
Moreover, the lack of detection of coronal lines in the optical spectrum of the Teacup could be indicating that this coronal region 
is more reddened than the NLR. Using the SDSS spectrum presented in 
\citet{Villar14} we measured [Fe VII]$\lambda$6087\AA/[O I]$\lambda$6300\AA=0.037$\pm$0.010, which is well below the range of
$\sim$0.6-5.7 reported by \citet{Rodriguez06b} for Seyfert galaxies with strong coronal lines detected in their NIR spectra.

Finally, in the case of the He I$\lambda$2.060 and the molecular lines, single Gaussians with FWHM$\sim$400--480 km~s$^{-1}$
were sufficient to reproduce the profiles (see Table \ref{tab1}). However, whilst neither the He I line nor the narrow core of the 
hydrogen recombination lines are shifted from the systemic velocity (as measured from the narrow component of Pa$\alpha$), 
all the H$_2$ lines are systematically 
blueshifted, with V$_s$=-51$\pm$32 km~s$^{-1}$ on average. This could be a first 
indication of a molecular outflow in the Teacup (see e.g. \citealt{Muller16} and references therein), although deeper 
observations are required to confirm it. 
It should also be noted that the use of the narrow component of Pa$\alpha$ as a tracer of the systemic velocity is affected by uncertainties
\citep{Villar14,Muller16}. Accurate measurements of the systemic velocity are needed to understand the kinematic behaviour of the 
H$_2$ lines.

\subsubsection{Emission line diagnostics}
\label{emission}

To determine the degree of obscuration of the nuclear region of the Teacup we calculated the narrow and broad 
Pa$\alpha$/Br$\gamma$ ratios in the two apertures considered here. 
By comparing them (see Table \ref{tab2}) with the theoretical value of 12.2 \citep{Hummer87} 
we can determine the 
optical and infrared extinction (A$_V$ and A$_K$ respectively) by using the parametrization A$_{\lambda}\propto\lambda^{-1.75}$
\citep{Draine89}. For the narrow component, we measure a maximum value of A$_V$=2.8$\pm$1.4 mag in the nucleus of the Teacup 
($\sim$830 pc diameter). In the large aperture ($\sim$2 kpc diameter) the level of obscuration decreases to 
0.55$\pm$1.14 mag. 

Following the same procedure, from the ratio of the broad components we measure A$_V$=3.5$\pm$1.9 mag 
in the nucleus of the Teacup and A$_V$=1.5$\pm$1.7 mag in the large aperture. The latter value of the extinction is consistent 
with A$_V$=1.9$\pm$0.3 mag reported by \citet{Villar14} for the nuclear outflow as measured from the SDSS spectrum. 

\begin{table}
\centering
\begin{tabular}{lcc}
\hline
\hline
 &\multicolumn{1}{c}{Nuclear spectrum} & \multicolumn{1}{c}{Aperture 1.25\arcsec} \\
\hline
Pa$\alpha^n$/Br$\gamma^n$       & 11.42$\pm$0.39  &    12.04$\pm$0.33  \\
A$_V^n$ (mag)                   & 2.77$\pm$1.39   &    0.55$\pm$1.14   \\
A$_K^n$ (mag)                   & 0.24$\pm$0.12   &    0.05$\pm$0.10   \\
Pa$\alpha^b$/Br$\gamma^b$       & 11.23$\pm$0.52  &    11.76$\pm$0.50  \\
A$_V^b$ (mag)                   & 3.46$\pm$1.90   &    1.54$\pm$1.73   \\
A$_K^b$ (mag)                   & 0.31$\pm$0.17   &    0.14$\pm$0.15   \\
\hline
H$_2$1-0S(1)/Br$\gamma^n$       & 0.61$\pm$0.40   &    0.94$\pm$0.34   \\

M$_{H_2}$ (10$^3 M_{\sun}$)     & 3.03$\pm$0.83   &    10.3$\pm$2.5   \\
M$_{cold}$ (10$^9 M_{\sun}$)    & 2.18$\pm$0.59   &    7.39$\pm$1.84   \\
\hline	   					 				 							        																																													  
\end{tabular}	
\caption{Emission-line ratios and derived properties measured from the narrow and broad components of the lines in the spectra 
extracted in the two apertures considered here (0.5\arcsec~and 1.25\arcsec~diameter, or equivalently
$\sim$0.8 and 2 kpc).
\label{tab2}}
\end{table}

The ratio H$_2$1-0S(1)/Br$\gamma$ can be used to disentangle the dominant excitation mechanism of the gas (see Table \ref{tab2}). 
In the two apertures considered we measure values in the range 0.6--0.9, consistent with AGN
photoinization (this ratio is lower than 0.6 for starburst galaxies and higher for LINERs; \citealt{Mazzalay13}). Although we
observe a tendency in 1-0S(1)/Br$\gamma$ to increase with the aperture, the values are consistent within the errors.

\begin{figure}
\includegraphics[width=6.5cm,angle=90]{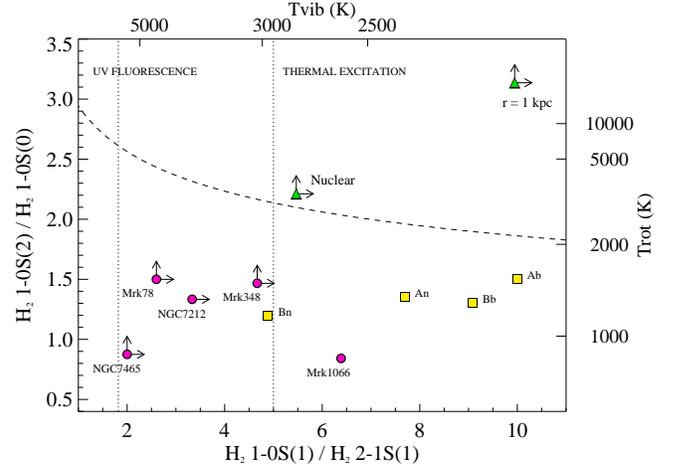}
\caption{Molecular line ratios measured in the two apertures considered here (0.5\arcsec and 1.25\arcsec; green triangles). 
Pink circles 
are the ratios measured from the nuclear spectra of the Seyfert 2 galaxies studied in \citet{Ramos09}. Yellow squares correspond to
the ratios derived for the broad and narrow components of regions A and B of the luminous IR galaxy (LIRG) and gas-rich merger 
NGC\,3256 \citep{Emonts14}. The dashed line indicates the locus of T$_{vib}$=T$_{rot}$. Vertical dotted lines delimitate the 
regions of ``thermal'' and ``non-thermal'' excitation from \citet{Mouri94}.}
\label{fig4}
\end{figure}

As can be seen from Figure \ref{fig2}, we detect several H$_2$ lines in the nuclear spectrum of the Teacup. In AGN, the lowest
vibrational levels (v=1) of H$_2$ tend to be thermalized (i.e. excited by shocks and/or X-ray illumination;
\citealt{Hollenbach89,Maloney96}), while higher level transitions are populated due to non-thermal processes such as UV 
fluorescence \citep{Black87}. The 1-0S(1)/2-1S(1) line ratio is an excellent discriminator between thermal and non-thermal 
processes. It is $\le$2 in gas excited by UV fluorescence and $\ge$5 in thermally-dominated gas \citep{Mouri94}. 
At the same time, the 1-0S(2)/1-0S(0) line ratio is sensitive to the strength of the incident radiation. 
In Figure \ref{fig4} we show the line ratios that we have measured in the
two apertures considered here. For comparison, we also plot the nuclear ratios of the five Seyfert 2 galaxies studied 
in \citet{Ramos09} and those measured for the broad and narrow component of the lines in regions A and B of the gas-rich merger and
luminous IR galaxy NGC\,3256 \citep{Emonts14}.

We find 1-0S(1)/2-1S(1) ratios $\ge$5, consistent with thermal excitation. Indeed, 
it can be seen from Figure \ref{fig4} that the position of the Teacup ratios in the diagram is very different from those 
measured in Seyfert 2 galaxies. On the other hand, the lower limits on the 1-0S(1)/2-1S(1) line ratio are consistent with the values 
reported by \citet{Emonts14} for regions A and B of the LIRG NGC\, 3256, but the 1-0S(2)/1-0S(0) values are significantly higher. 
This implies that although the molecular gas in the Teacup and in NGC\,3256 is thermally excited, 
the strength of the incident radiation is not the same. This is expected considering that the presence of nuclear activity has not
been confirmed yet in NGC\,3256 \citep{Emonts14}, while the Teacup hosts a very luminous AGN.

Using the two line ratios mentioned above we can derive the rotational and vibrational temperatures of the gas
following \citet{Reunanen02}. Although we only have upper limits, in the case of the nuclear spectrum T$_{vib}\simeq
T_{rot}\ga 3000$ K, which is characteristic of thermally excited gas. On the other hand, for the large aperture 
T$_{vib}<<T_{rot}$, indicating that gas excitation is more complex than local thermal equilibrium (LTE) conditions (T$_{vib}=T_{rot}$). 


Finally, we can use the H$_2$1-0S(1) luminosity to estimate the amount of molecular gas present in the nucleus of the Teacup.
Following \citet{Mazzalay13}, the relation between the line flux and the warm molecular gas mass is

\begin{equation}
M_{H_2}\simeq 5.0875 \times 10^{13} (\frac{D}{Mpc})^2 (\frac{F_{1-0S(1)}}{erg~s^{-1} cm^{-2}})10^{0.4A_K},
\label{eq1}
\end{equation}

where D=387 Mpc and A$_K$ is the extinction reported in Table \ref{tab2}. In the same Table we show M$_{H_2}$ measured in 
the two apertures considered, as well as the masses of cold molecular gas (M$_{cold}$) calculated by assuming a cold--to--warm 
mass ratio M$_{cold}$/M$_{H_2}\simeq0.7\times10^{6}$. This ratio was derived observationally by \citet{Mazzalay13} by comparing values 
of M$_{cold}$ obtained from CO
observations and H$_2$ luminosities for a large number of galaxies covering a wide range of luminosities, morphological types and 
nuclear activity. A similar ratio was reported by \citet{Dale05} for a large sample of active and star-forming galaxies 
(M$_{cold}$/M$_{H_2}\simeq10^{5-7}$).


In the nucleus of the Teacup we measure 
M$_{H_2}=(3.0\pm0.8)\times10^3 M_{\sun}$ and M$_{cold}=(2.2\pm0.6)\times10^9 M_{\sun}$.
If instead of looking at the 
inner $\sim$830 pc of the Teacup we measure the molecular gas content in the inner 2 kpc, 
M$_{H_2}=(1.0\pm0.2)\times10^4 M_{\sun}$ and M$_{cold}=(7.4\pm1.8)\times10^9 M_{\sun}$. 
These cold gas masses are similar to those 
reported by \citet{Villar13} for a sample of 10 QSO2s at z$\sim$0.2--0.3 with CO measurements. 



\subsection{Ionized, coronal and molecular emission-line maps}
\label{extended}

In the previous sections we studied the spectra of the nuclear region of the Teacup. 
Here we take advantage of the spatial information provided by SINFONI and study the flux distribution and kinematics 
of Pa$\alpha$, [Si VI] and H$_2$1-0S(1). These three emission lines are the highest S/N representatives of the ionized, coronal and
molecular phases of the gas respectively. 

\begin{figure*}
\includegraphics[width=18.0cm]{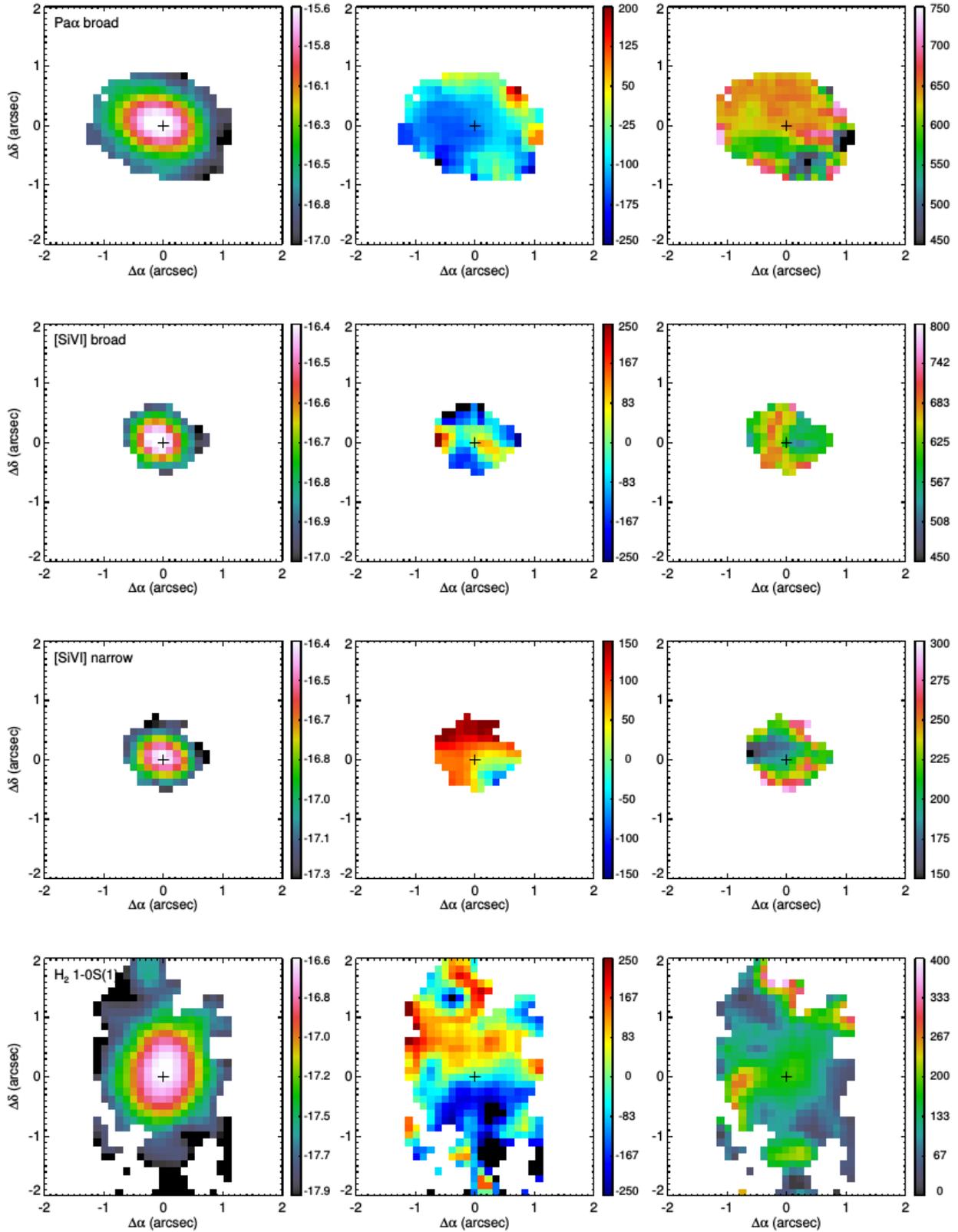}
\caption{Flux, velocity and velocity dispersion maps of the broad Pa$\alpha$, broad and narrow [Si VI], and H$_2$1-0S(1)
lines. The continuum-subtracted flux maps are in
logaritmic scale and units are erg~$s^{-1}$~cm$^{-2}$. The velocity and dispersion maps are in km~s$^{-1}$, and velocities are 
relative to the narrow core of Pa$\alpha$.
The maps have been smoothed using a 2-spaxels boxcar. The brightest spaxel in the ionized gas emission is indicated with a cross.
1\arcsec~corresponds to 1.6 kpc. North is up and East to the left.}
\label{fig5}
\end{figure*}

Pa$\alpha$ is the only line that we detect in the extended emission-line structures (i.e. the NE bubble and the SW fan; see Section \ref{bubbles}). 
We note, however, that the SINFONI FOV (9.2\arcsec$\times$8.7\arcsec) does not cover the full extent of the NE bubble, as shown 
in Figure \ref{fig1}. 
In the case of the nuclear Pa$\alpha$ and [Si VI] emission we needed two Gaussians to reproduce the observed 
line profiles and obtain corresponding flux, velocity and velocity dispersion ($\sigma$) maps. 
For the H$_2$1-0S(1) emission line (hereafter H$_2$) a single Gaussian was sufficient. Since the [Si VI] and H$_2$1-0S(3)
lines are blended, we fixed the FWHM of the latter emission line to match that of H$_2$1-0S(1) and get reliable fits for the [Si VI].
In Figure \ref{fig5}
we show the flux, velocity and $\sigma$ maps of the broad Pa$\alpha$, broad and narrow [Si VI] components, and H$_2$. 
These maps correspond to a 4\arcsec$\times$4\arcsec~(6.4$\times$6.4 kpc$^2$) FOV. 

Using the flux maps we can estimate the projected
sizes of each emission-line region. In Table \ref{tab3} we show the observed and seeing-deconvolved FWHMs 
along the major axis in each case. We note that the intrinsic, deconvolved sizes for the central distribution of the emission lines considered here
are approximate sizes. They have been obtained by applying the standard Gaussian deconvolution 
method based on the well-known quadrature relation FWHM$_{int}^2$ = FWHM$_{obs}^2$ - FWHM$_{seeing}^2$. All the emission-line maps shown
in Figure \ref{fig5} are resolved except the narrow [Si VI] maps.

\begin{table}
\centering
\begin{tabular}{lcccc}
\hline
\hline
Line & FWHM  &  \multicolumn{2}{c}{Seeing-decnv. FWHM} & PA  \\
& (arcsec)   &  (arcsec)     & (kpc) & (deg) \\
\hline
Pa$\alpha$ (b)  & 0.76$\pm$0.06 &  0.55$\pm$0.09  & 0.88$\pm$0.14 & 75.3$\pm$0.2 \\
$[Si~VI]$ (b)	& 0.75$\pm$0.06 &  0.54$\pm$0.09  & 0.86$\pm$0.14 & 72.4$\pm$0.3 \\
$[Si~VI]$ (n)	& 0.60$\pm$0.06 &  \dots  & \dots & 73.9$\pm$0.4 \\
H$_{2}$ 1-0S(1)	& 1.00$\pm$0.06 &  0.85$\pm$0.09  & 1.36$\pm$0.14 & -7.7$\pm$0.5 \\
\hline	   				 			    																																				      
\end{tabular}	
\caption{Radial sizes along the major axis and corresponding PAs of the broad Pa$\alpha$, broad and narrow [Si VI], and H$_2$ emitting regions as
measured from the flux maps shown in Figure \ref{fig5}. Column 2 corresponds to observed FHWMs and columns 3 and 4 to seeing-deconvolved FWHMs.}
\label{tab3}
\end{table}

The broad Pa$\alpha$ and broad [Si VI] flux maps 
show similar morphologies, and are both elongated roughly in the same direction 
(PA$\sim$72\degr--75\degr). We note that this elongation does not coincide 
with the orientation of the seeing major axis (PA=88.3$\pm$0.1\degr), as measured from the standard star. 
In the case of the broad [Si VI] flux map, two peaks are observed, 
separated by $\sim$0.3\arcsec~with a PA$\sim$70\degr~(although it is difficult to distinguish them in the [Si VI] flux map shown in 
Figure \ref{fig5}, they are evident when displayed in contours). 
One of these flux peaks coincides with the AGN position, as traced by the 
maximum of the Pa$\alpha$ emission, and the other could be the coronal counterpart of the compact HR-B region detected by 
\citealt{Harrison15} in the high-angular resolution VLA radio maps. The HR-B compact region is located 0.5\arcsec~northeast of the 
AGN position, with PA$\sim$60\degr. 
As expected for outflow-related components, the maps of the broad Pa$\alpha$ and broad [Si VI] emission are 
dominated by blueshifted velocity components, reaching maximum values of -250 km~s$^{-1}$. 

The narrow [Si VI] emission-line maps (third row of Figure \ref{fig5}) are similar to
those of the broad component, with the exception of the velocity map. The NE part of the emission 
is redshifted and the SW blueshifted, with maximum velocities of $\pm$150 km~s$^{-1}$. 
From the narrow [Si VI] flux map we can set constraints on the size of the coronal line region (CLR) of the Teacup. 
The observed FWHM that we measure is consistent with the seeing FWHM within the errors (see Table \ref{tab3}). Therefore 
the size of the CLR is formally unresolved and we can estimate an upper limit as
[(FWHM$_{seeing}$+3$\sigma_{seeing}$)$^2$-FWHM$_{seeing}^2]^{0.5}$ = 0.47\arcsec$\sim$746 pc.

 

The H$_2$ maps (bottom panels of Figure \ref{fig5}) are completely different to those of the ionized and coronal lines. 
They are elongated almost in the N--S direction (PA=-7.7$\pm0.5\degr$), which is roughly perpendicular to the
orientation of the Pa$\alpha$ and [Si VI] maps. The seeing-deconvolved FWHM along the major axis is 1.4$\pm$0.2 kpc
(see Table \ref{tab3}).
The global H$_2$ velocity field seems to be dominated by rotation, with maximum velocities of $\pm$250 km~s$^{-1}$.
Deviations from this pattern are identified at some spatial locations, but they correspond to spaxels where the 
signal-to-noise is lower.  
The $\sigma$ values are at a maximum in the central region (FWHM$\sim$450 km~s$^{-1}$) and 
decrease toward the edges (FWHM$\sim$150 km~s$^{-1}$). 


\subsubsection{Extent of the nuclear outflow}

\begin{figure*}
\includegraphics[width=13cm]{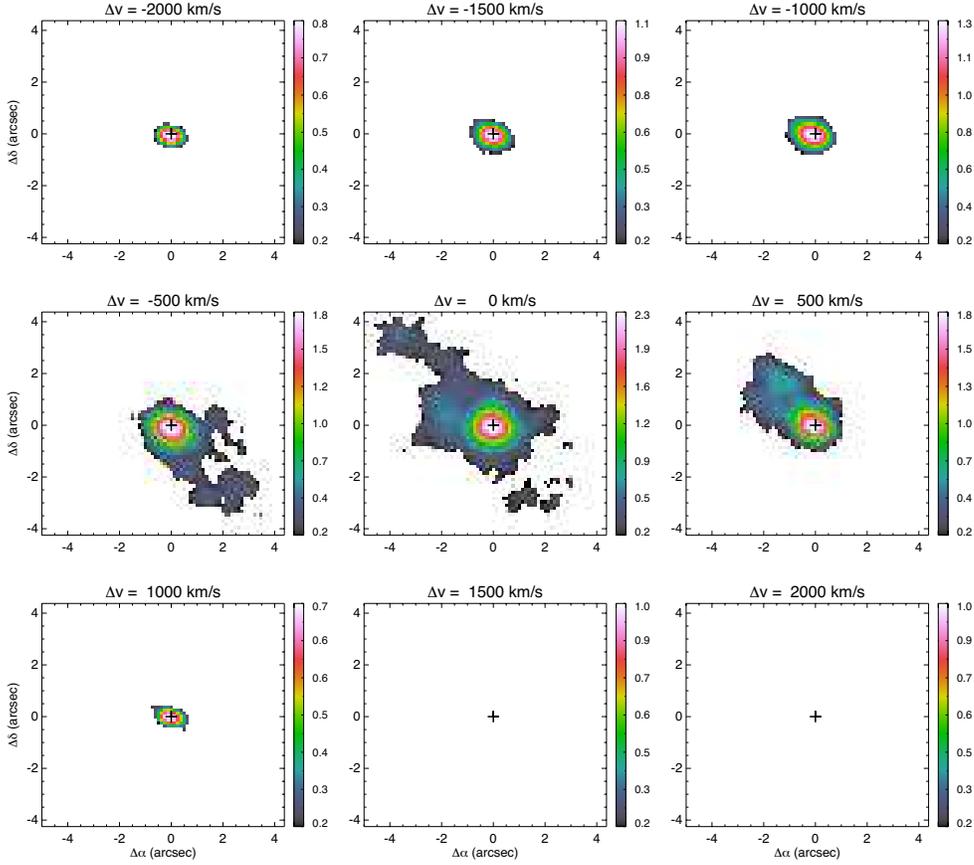}
\caption{Continuum-subtracted Pa$\alpha$ flux maps extracted in consecutive velocity intervals of 500 km~s$^{-1}$, 
centred at the maximum of the line profile in the central spaxel (see Figure \ref{fig7}). The maps are in
units of Log ($\sigma$) and have been smoothed using a 2-spaxels boxcar.
North is up and East to the left.}
\label{fig6}
\end{figure*}
\begin{figure}
\centering
\includegraphics[width=8cm]{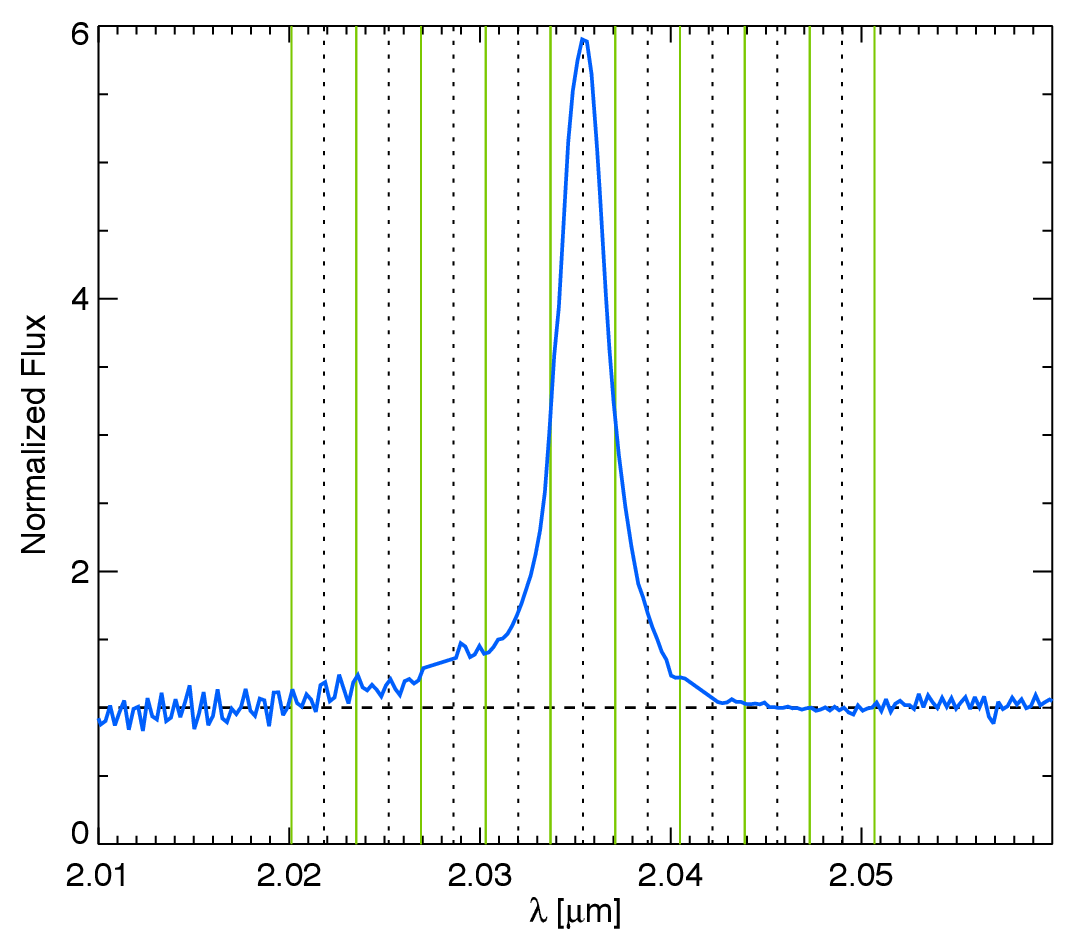}
\caption{Pa$\alpha$ line profile as extracted from the central spaxel. The continuum has been normalized (dashed line).
The solid-green vertical lines indicate the velocity bins (from -2000 
to 2000 km~s$^{-1}$ in steps of 500 km~s$^{-1}$) used for extracting the nine flux maps shown in Figure \ref{fig6}.}
\label{fig7}
\end{figure}

In Figure \ref{fig6}
we show Pa$\alpha$ flux maps extracted in consecutive velocity intervals of 500 km~s$^{-1}$, centred at the 
maximum of the line profile in the central spaxel (see Figure \ref{fig7}). 
These velocity cuts allow us to characterize the extent and orientation of the Pa$\alpha$ emission in the core and the wings 
of the line. 


In the central panel we see the Pa$\alpha$ emission corresponding to the core of the line, which we have assumed as
systemic redshift. From this map we can distinguish the nuclear emission (inner $\sim$2\arcsec) and the extent of the 
emission-line features covered by the
SINFONI FOV (the NE bubble and the SW fan). The right and left middle panels correspond to the velocity bins 
centred at $\pm$500 km~s$^{-1}$, so the Pa$\alpha$ wings dominate the emission. The nuclear emission 
appears elongated in the two panels and on larger scales, depending on whether we are sampling the blue or the red wing of Pa$\alpha$, 
we see the SW fan or the NE bubble, respectively.

The blueshifted broad component that we identified in the nuclear spectrum of the Teacup is sampled by the four negative velocity
bins (see Figure \ref{fig7}). We fitted a Gaussian model to the nuclear region detected in the corresponding flux maps (middle left
and top panels in Figure \ref{fig6}) and measured seeing-deconvolved FWHMs ranging from 1 to 1.1 kpc along the major axis, with
PA=70\degr--75\degr.
We measure a similar extension and orientation of the outflow when we sample the red
wing of Pa$\alpha$ (middle right and bottom left panels in Figure \ref{fig6}). This indicates that we can
trace the nuclear outflow up to 1000 km~s$^{-1}$ beyond the core of Pa$\alpha$. It is noteworthy that 
the elongation of the nuclear Pa$\alpha$ emission that we see at the highest velocities is less obvious
at $\Delta$v=0 km~s$^{-1}$ (i.e. at the core of the line), confirming that the nuclear outflow is extended. 

Therefore from the analysis of Figure \ref{fig6} we conclude that the nuclear ionized outflow is resolved. The 
observed maximum extension, which corresponds to the -1000 km~s$^{-1}$ velocity bin, 
is FWHM=1.37$\pm$0.06\arcsec~along the 73.9$\pm$0.2\degr~axis. As noted in Section \ref{observations}, the
standard star that was used to characterize the seeing is slightly elongated along PA=88.3$\pm$0.1\degr, 
with a maximum extension of 0.58\arcsec. 
Even considering the maximum value of the seeing, the nuclear ionized outflow is resolved, with a seeing-deconvolved 
FWHM=0.68$\pm$0.09\arcsec~(1.1$\pm$0.1 kpc). 

We cannot repeat the same exercise for the [Si VI] emission line because its blue wing is blended with H$_2$1-0S(3) and it has 
lower signal-to-noise than Pa$\alpha$. However, if we fit a
Gaussian component to the broad [Si VI] flux map shown in Figure \ref{fig5}, we measure a FWHM=0.75$\pm$0.06\arcsec, which 
corresponds to a seeing-deconvolved 
FWHM=0.54$\pm$0.09\arcsec~(0.9$\pm$0.1 kpc) along the major axis (PA=72.4$\pm$0.3\degr). Therefore, the coronal outflow is also 
resolved in the same direction as the ionized outflow.

\subsubsection{Large-scale Pa$\alpha$ emission}
\label{bubbles}

\begin{figure*}
\centering
\includegraphics[width=18cm]{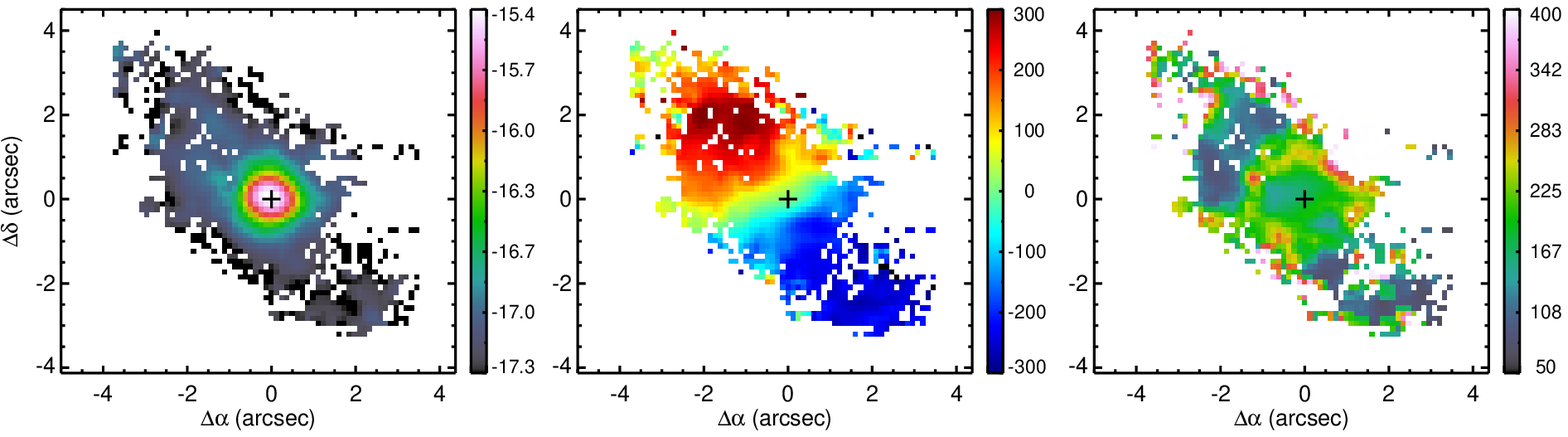}
\caption{Same as in Figure \ref{fig5} but for the narrow component of Pa$\alpha$ and in a larger FOV of 
9.5\arcsec$\times$9\arcsec~(15.1$\times$14.3 kpc$^2$).}
\label{fig8}
\end{figure*}

In Figure \ref{fig8}
we show the flux, velocity and dispersion maps of the narrow component of Pa$\alpha$ in a 9.5\arcsec$\times$9\arcsec~(15.1$\times$14.3
kpc$^2$) FOV.  
The flux map resembles the morphology of the [O III] and H$\alpha$ 
HST images \citep{Keel15}. The bulk of the Pa$\alpha$ emission is dominated by the nuclear component, 
and beyond we detect part of the NE bubble (radial size of $\sim$9.5 kpc), and the SW fan ($\sim$7 kpc). 
The velocity map (central panel of Figure \ref{fig8}) is similar to the [O III] 
velocity map reported by \citealt{Harrison15}, although it reveals that the kinematics of the gas in the bubble and fan are 
different from those of the NLR. Whilst the central $\sim$2\arcsec~show a smooth velocity field, probably coincident with the rotation 
pattern of the galaxy, beyond this there is an abrupt change in velocity.
The NE bubble is redshifted with maximum velocities of 300 
km~s$^{-1}$ relative to systemic, and the SW fan is blueshifted with V$_{max}$=-300 km~s$^{-1}$. 
The dispersion map also shows a practically
constant line width within the central 2\arcsec~(FWHM$\sim$450-500 km~s$^{-1}$) and smaller values in the bubble and fan 
(FWHM$\sim$200--300 km~s$^{-1}$). We note that the FWHM$>$500 km~s$^{-1}$ (i.e. $\sigma\ga 210$ km~s$^{-1}$) that we measure 
around the nuclear region ($\sim$1\arcsec~radius) corresponds to the transition zone where we no longer need to
fit a broad component to reproduce the Pa$\alpha$ profiles. 

\begin{figure*}
\centering
{\par\includegraphics[width=6.1cm,angle=90]{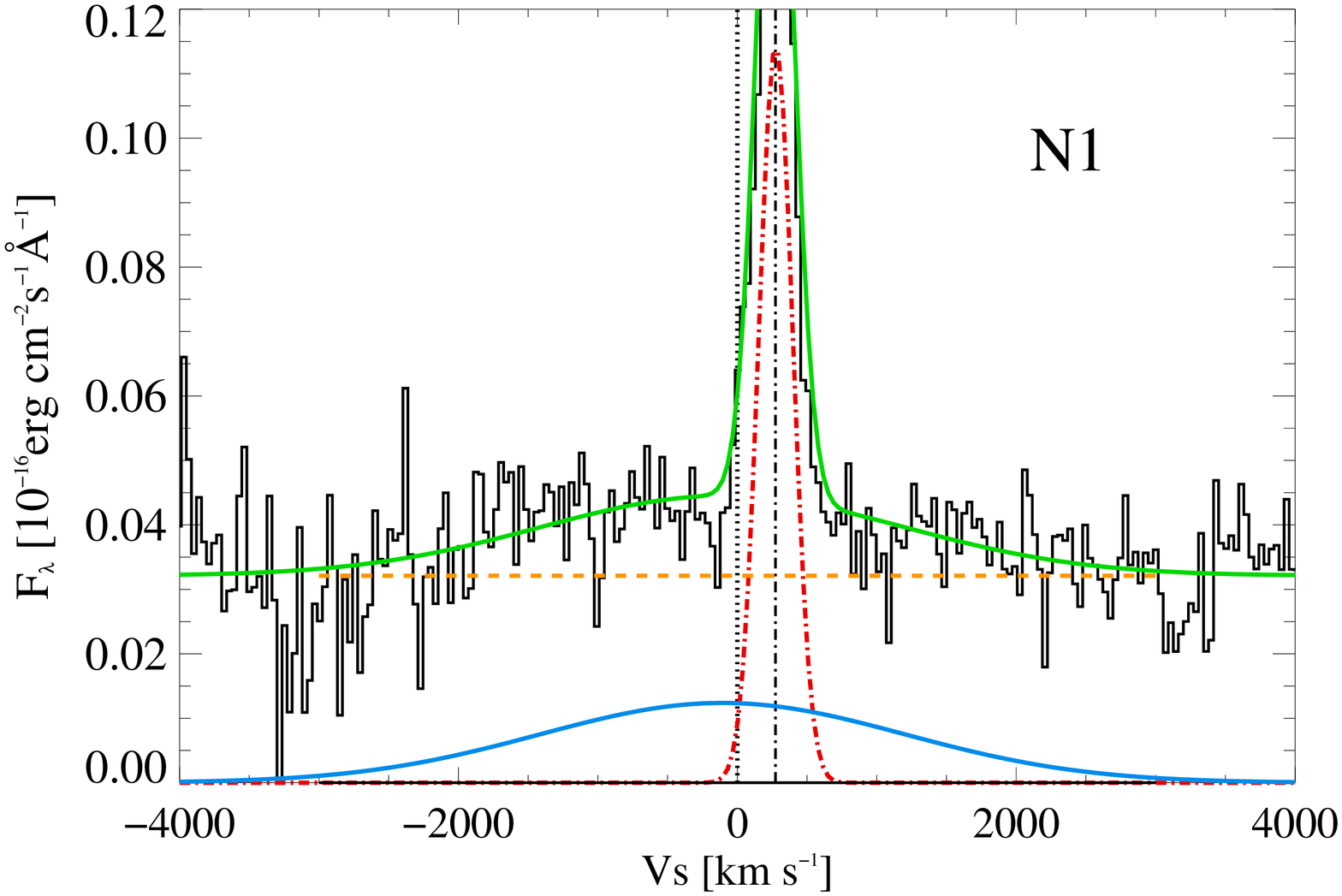}
\includegraphics[width=6.1cm,angle=90]{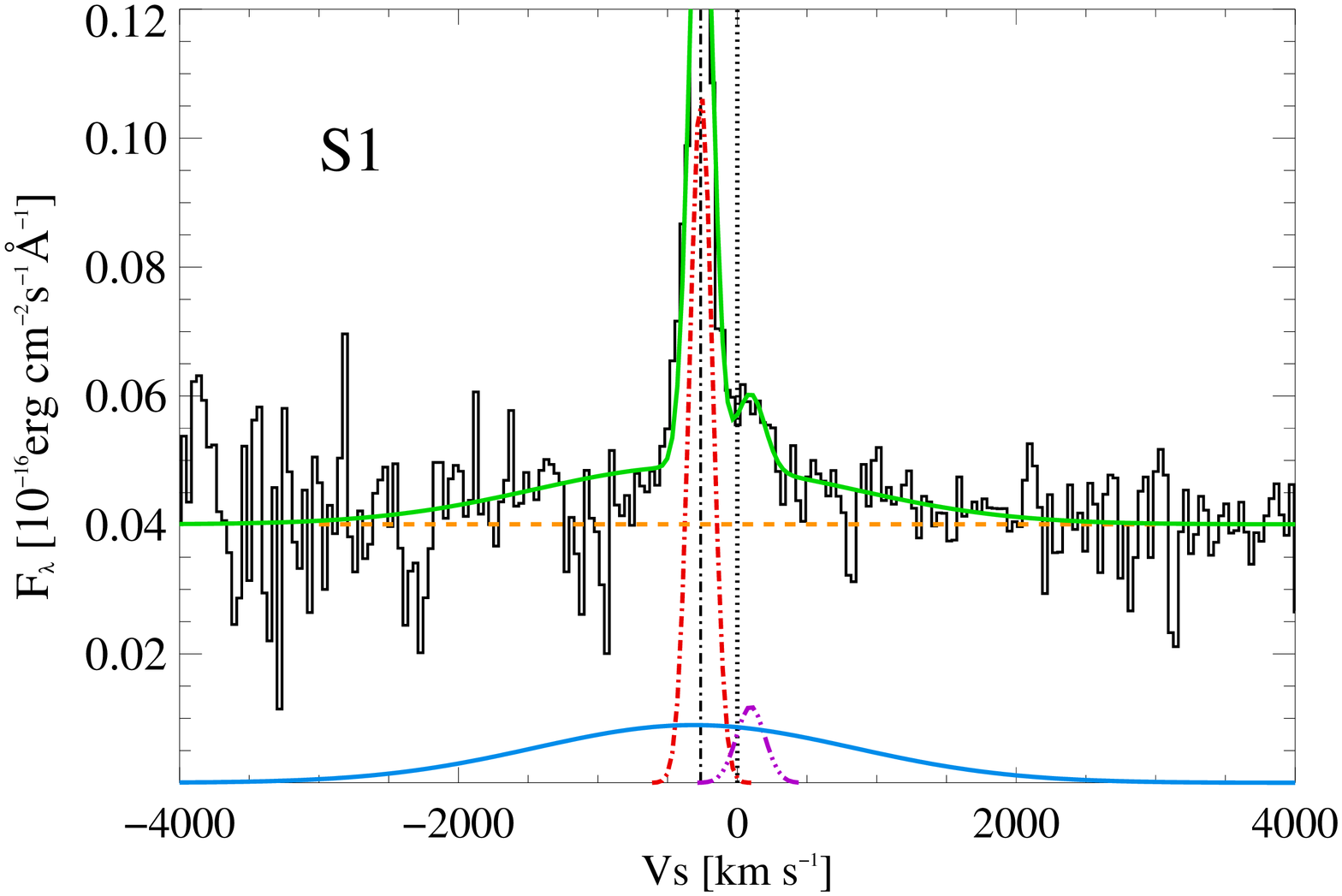}
\includegraphics[width=6.1cm,angle=90]{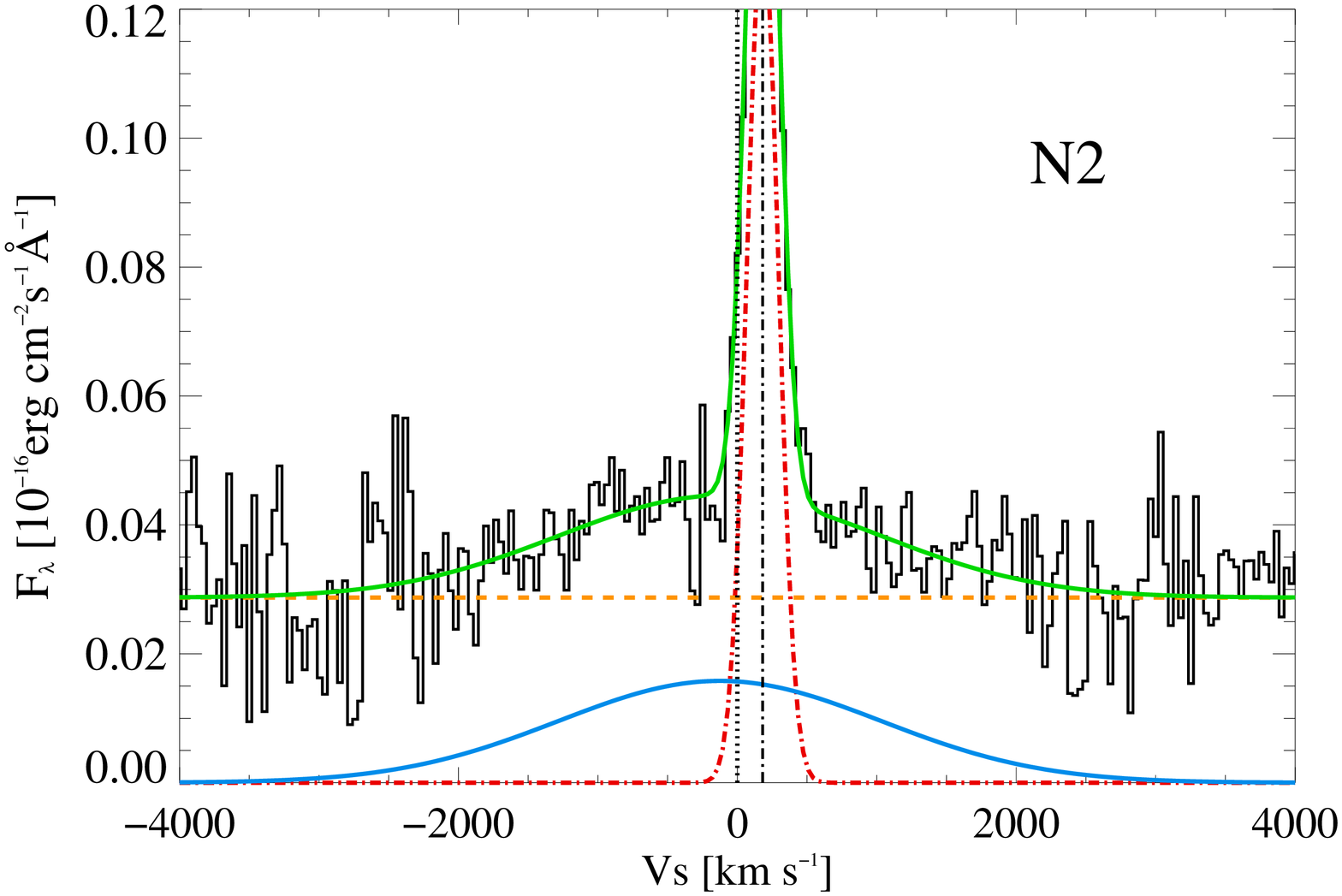}
\includegraphics[width=6.1cm,angle=90]{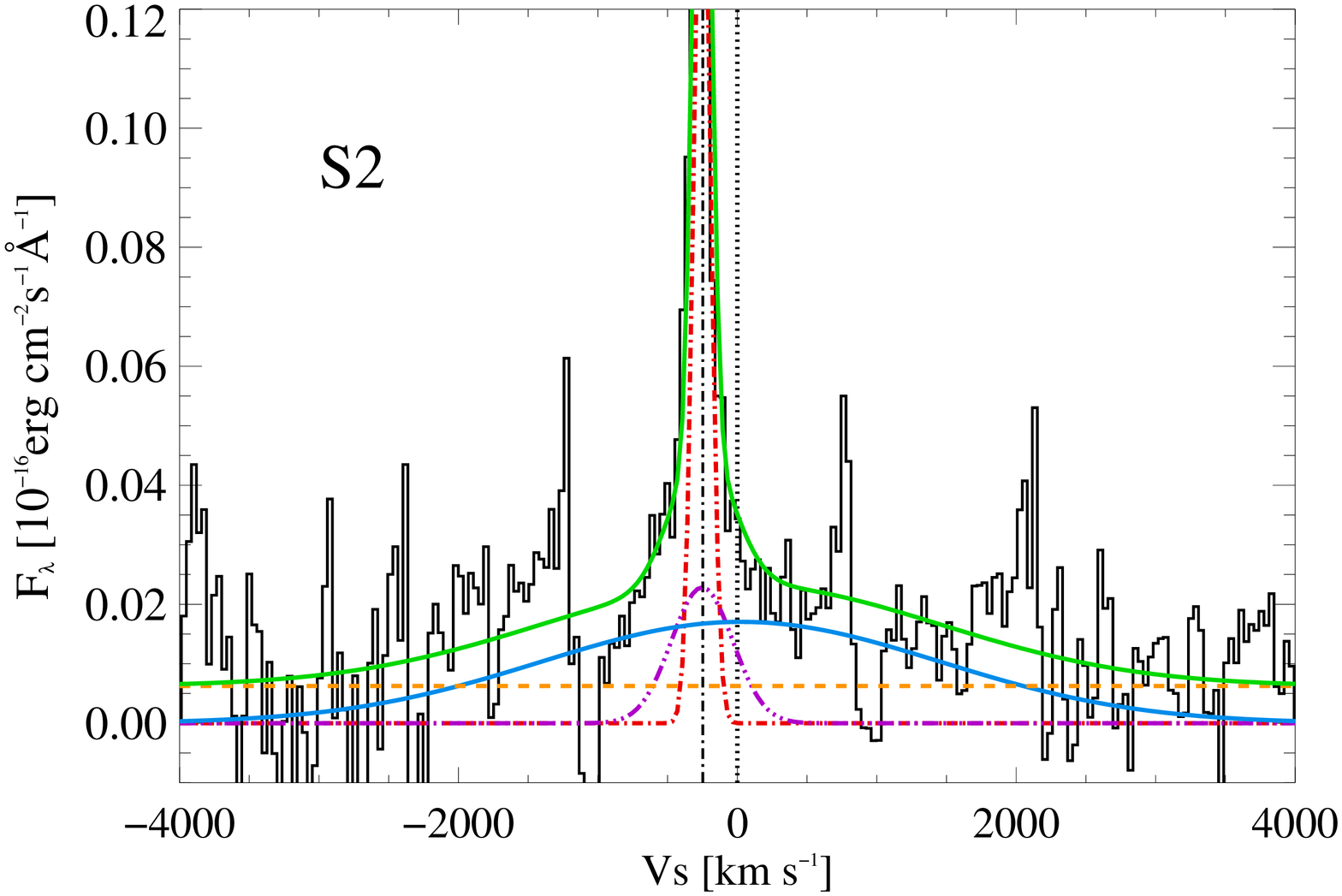}
\includegraphics[width=6.1cm,angle=90]{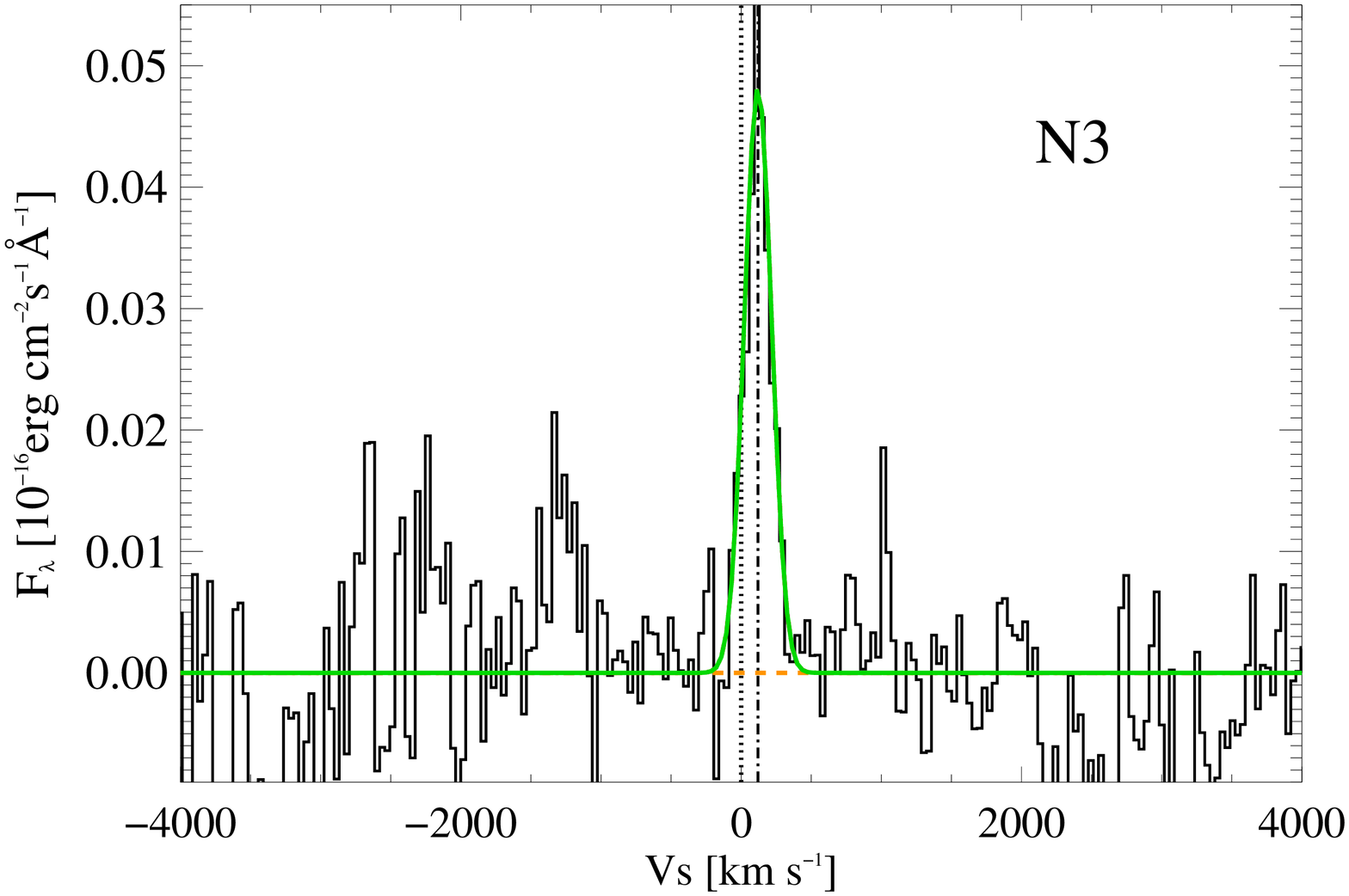}
\includegraphics[width=6.1cm,angle=90]{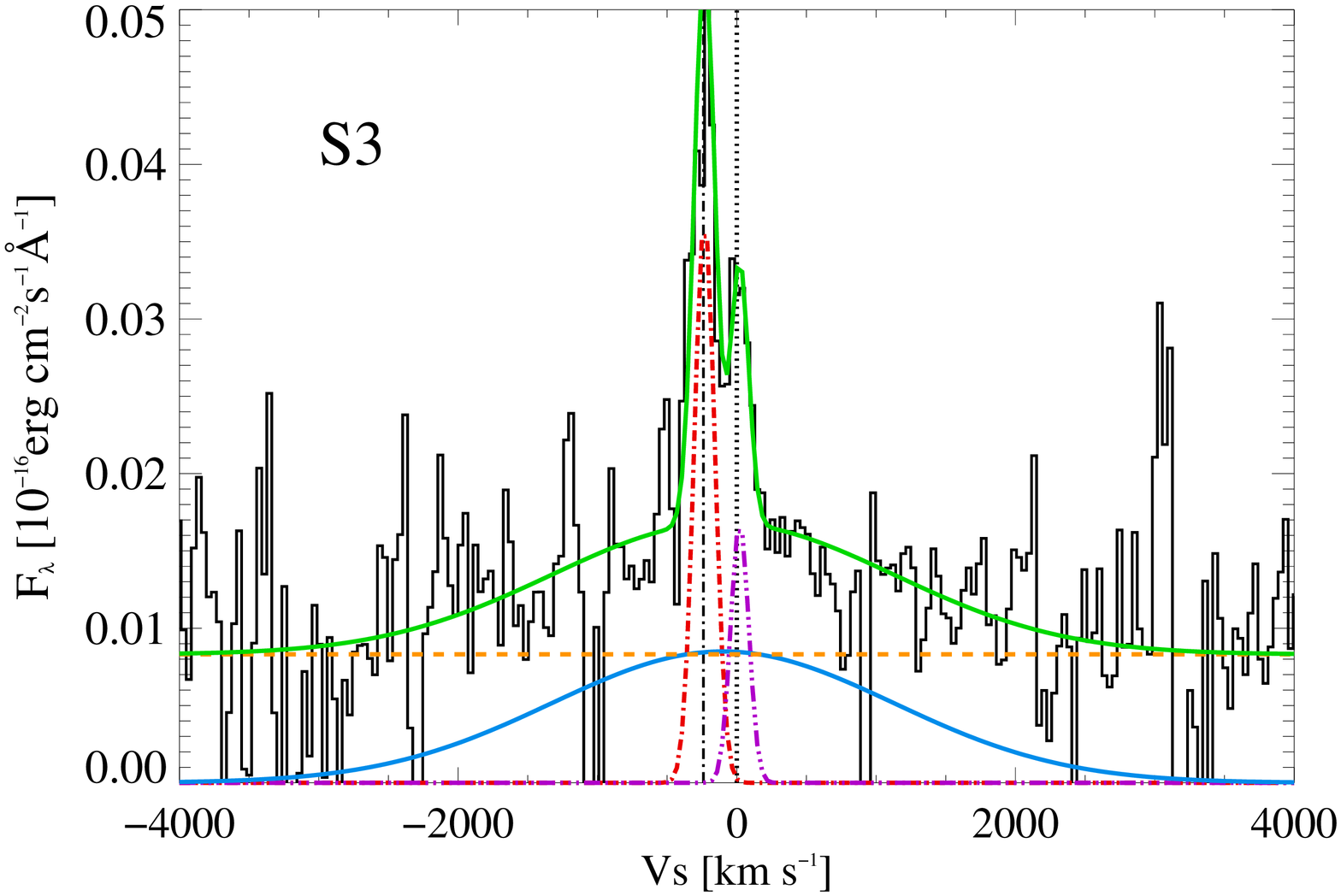}\par}
\caption{Pa$\alpha$ line profiles detected at the different spatial locations indicated in Figure \ref{fig1} and Table \ref{tab4}.  
N1, N2, and N3 spectra correspond to the NE bubble and S1, S2, and S3 to the SW fan. 
Corresponding fits and continua are shown as green solid and orange dashed lines. 
Dot-dashed red Gaussians correspond to the narrow Pa$\alpha$ component and solid
purple Gaussians to the additional Pa$\alpha$ components described in Table \ref{tab4}. 
The very broad components tentatively detected in all the regions considered except N3 are shown as solid blue Gaussians.
The vertical dotted line indicates zero velocity as set from the 
nuclear narrow component of Pa$\alpha$, and the dot-dashed line is V$_s$ measured for each line.}
\label{fig9}
\end{figure*}

To study the properties of the extended Pa$\alpha$ emission of the Teacup in more detail, we extracted spectra from
different spatial locations in the bubble and fan. In Figure \ref{fig1} we indicate the positions and apertures chosen 
for extracting the spectra in the galaxy nucleus, bubble (N1, N2, and N3) and fan (S1, S2, and S3). In 
Table \ref{tab4} we report the different apertures chosen in each case and the radial distances from the nucleus. 
We show the six spectra extracted in the bubble and the fan in Figure \ref{fig9}. We used {\sc dipso} to fit the emission lines 
using Gaussian profiles, and in Table \ref{tab4} we report the corresponding FWHMs, V$_s$ and line intesities.
In the case of the bubble, single Gaussian components of FWHM$\sim$300 km~s$^{-1}$ were sufficient to reproduce the narrow 
line profiles in the N1 and N2 spectra, extracted at $\sim$3
kpc from the nucleus. For the more distant N3 region ($\sim$8 kpc from the nucleus) the FWHM is narrower 
(FWHM$\sim$200 km~s$^{-1}$). We note that, as we already knew from the Pa$\alpha$ velocity map shown in Figure \ref{fig8}, 
these narrow Pa$\alpha$ lines are significantly redshifted relative to the systemic velocity
(V$_s$$\sim$120--270 km~s$^{-1}$; see Figure \ref{fig9}).

\begin{table*}
\small
\centering
\begin{tabular}{lccccccccccc}
\hline
\hline
Region & \multicolumn{2}{c}{Distance} & Aperture & \multicolumn{3}{c}{Pa$\alpha$ narrow component} & \multicolumn{3}{c}{Pa$\alpha$ additional component} \\
&           \multicolumn{2}{c}{to nucleus}  &diameter  &  FWHM & V$_s$ & Flux $\times~10^{16}$  &  FWHM & V$_s$ & Flux $\times~10^{16}$ \\
& (arcsec) & (kpc) & (arcsec) & (km~s$^{-1}$) & (km~s$^{-1}$) & (ergs~cm$^{-2}$~s$^{-1}$)  & (km~s$^{-1}$) & (km~s$^{-1}$) & (ergs~cm$^{-2}$~s$^{-1}$) \\
\hline
r2     	    &   0.0 & 0.0    & 0.50\arcsec &  434$\pm$7      &   0$\pm$3   & 59.6$\pm$1.3   &   1794$\pm$93  & -234$\pm$36  & 36.4$\pm$1.5    \\
N1          &	2.1 & 3.4    & 0.75\arcsec &  305$\pm$35     & 274$\pm$17  & 2.77$\pm$0.10  &   \dots        &   \dots      &   \dots         \\
N2     	    &   1.7 & 2.8    & 0.75\arcsec &  285$\pm$13     & 182$\pm$10  & 3.05$\pm$0.11  &   \dots        &   \dots      &   \dots         \\
N3          &	5.1 & 8.1    & 0.50\arcsec &  211$\pm$43     & 120$\pm$19  & 0.80$\pm$0.12  &   \dots        &   \dots      &   \dots         \\
S1          &	1.8 & 2.9    & 0.75\arcsec &  191$\pm$14     &-263$\pm$10  & 1.68$\pm$0.10  &   307$\pm$116  &  116$\pm$42  &   0.40$\pm$0.12 \\
S2     	    &   3.5 & 5.6    & 1.00\arcsec &   93$\pm$28     &-248$\pm$13  & 1.40$\pm$0.43  &   511$\pm$245  & -300$\pm$41  &   1.62$\pm$0.35 \\
S3          &	2.0 & 3.2    & 0.50\arcsec &  205$\pm$41     &-239$\pm$16  & 0.63$\pm$0.09  &   182$\pm$67   &   43$\pm$26  &   0.30$\pm$0.08 \\
\hline	   					 			    																																				     
\end{tabular}	
\caption{Gaussian components fitted to the Pa$\alpha$ emission line profiles detected in the nuclear spectrum and in the six apertures considered here. 
We note that these fits do not include the very broad components shown as blue Gaussians in Figure \ref{fig9}.
FWHMs are corrected from instrumental broadening, and velocity shifts (V$_s$) are relative to the nuclear Pa$\alpha$ 
narrow component.}
\label{tab4}
\end{table*}

We also extracted three spectra mapping two bright regions within the fan (S1 and S2) and a bright knot
at $\sim$3 kpc towards the west (S3). Two Gaussians
were necessary to reproduce the narrow Pa$\alpha$ profiles detected in the three spectra (see right panels of Figure \ref{fig9}). First, 
the blueshifted (V$_s\sim$-250 km~s$^{-1}$) narrow components with FWHM=100--200 km~s$^{-1}$ already
shown in the velocity map (see Figure \ref{fig8}). Second, in the case of the S1 and S3 spectra we required an additional
component of FWHM$\sim$200-300 km~s$^{-1}$ redshifted by V$_s\sim$40--120 km~s$^{-1}$ from systemic.  
For the S2 spectrum we fitted a slightly broader component of FWHM$\sim$500 km~s$^{-1}$ whose V$_s$ is consistent
with that of the narrow component (V$_s\sim$-300 km~s$^{-1}$).

Finally, we report tentative detection of a very broad Pa$\alpha$ component 
of FWHM$\sim$3000 km~s$^{-1}$ in five of the regions considered here (shown as blue Gaussians in Figure \ref{fig9}). 
These broad components are blueshifted 
relative to systemic in the bubble and fan, with V$_s$ ranging between -60 and -310 km~s$^{-1}$.
Unfortunately, the combination of sky residuals and low signal--to--noise of the spectra 
prevents confirmation of these very broad Pa$\alpha$ components. For this reason the results from the fits
reported in Table \ref{tab4} do not include the broad components.


\section{Discussion}
\label{discussion}


\subsection{The nuclear outflow}


Based on the blueshifted broad lines detected in the nuclear K-band spectrum of the Teacup we confirm the presence 
of the nuclear ionized outflow previously 
reported by \citet{Villar14} and \citealt{Harrison15} using optical spectra, and we report the existence of its
coronal counterpart. Although coronal outflows are commonly detected in Seyfert galaxies 
(e.g. \citealt{Gelbord09,Davies14}), this is one of the first detections of coronal outflows 
in QSO2s. Another example is Mrk\,477, a QSO2 at z=0.037 with an [O III] luminosity of 3.3$\times$10$^{42}~erg~s^{-1}$ for 
which \citet{Villar15} reported a blueshifted component of FWHM=2460$\pm$340 km~s$^{-1}$ in the high-ionization line 
[Fe VII]$\lambda$6087 \AA~(IP=99.1 eV).

The FWHM of this blueshifted component is larger than those measured in the optical for the central 3\arcsec~of the Teacup 
using the SDSS spectrum (FWHM$\sim$1000 km~s$^{-1}$; \citealt{Villar14}) and GMOS/Gemini IFU data 
(maximum FWHM$\sim$1000 km~s$^{-1}$; \citealt{Harrison14}). This is consistent with the outflow being reddened, as  
first claimed by \citet{Villar14} and confirmed in this work, supporting the advantage of using NIR observations to 
trace AGN outflows closer to their origin. 

We measure V$_s$=-234$\pm$35 km~s$^{-1}$ 
for the nuclear broad Pa$\alpha$ component. This value is consistent with the velocities reported by \citet{Harrison14} for the
blueshifted component of the [O III] line measured from GMOS IFU data. They report a velocity offset of 
$\Delta$v=-150 km~s$^{-1}$ for the broad component in the galaxy-integrated spectrum, which corresponds to the central 
5\arcsec$\times$3.5\arcsec.
However, if we look at the velocity maps (Figure A14 in \citealt{Harrison14}), there is a strong velocity gradient around the 
position of the AGN, with a maximum velocity offset of -268 km~s$^{-1}$. The broad Pa$\alpha$ velocity map that we report here 
resembles the [O III] velocity field, showing maximum velocities of -250 km~s$^{-1}$.

On the other hand, \citet{Villar14} reported V$_s$=-70 km~s$^{-1}$ for the broad 
component of [O III] relative to the narrow core, but measured in an aperture of 3\arcsec~diameter ($\sim$5 kpc).
The broad Pa$\alpha$ maps, shown in the top middle panel of Figure \ref{fig5}, demonstrate that the SDSS fiber 
includes both blueshifted and redshifted components, which explains the smaller V$_s$ reported in \citet{Villar14}.


\citealt{Harrison15} extracted a 
spectrum at the position of the HR-B region detected in the VLA radio maps and measured a FWHM=720 km~s$^{-1}$ 
with an observed velocity of -740 km~s$^{-1}$ relative to the narrow component of [O III]. 
Our SINFONI data have very high S/N in the nuclear region and yet, we do not require such a high velocity
component to reproduce any of the profiles (see top middle panel of Figure \ref{fig5}). 
This could be due to 
the combined effect of reddening variation at different spatial locations, different aperture sizes and 
kinematic substructure within the outflow region.


Using the Pa$\alpha$ and [Si VI] flux maps we confirm that both the ionized and coronal nuclear outflows are extended. We 
derived radial sizes of 1.1$\pm$0.1 kpc and 0.9$\pm$0.1 kpc respectively, with PA=72\degr--74\degr. This PA is almost identical 
to the radio jet orientation measured from the 1.4 GHz FIRST radio maps (PA$\sim$77\degr; \citealt{Harrison14}). 
This could be indicating, as first suggested by \citealt{Harrison15}, that the interaction
between the radio jet and the galaxy interstellar medium may have triggered and accelerated the nuclear outflow. 
The Teacup is then likely an example of radio jets driving outflows in a radio-quiet AGN. 
HST studies of Seyfert galaxies with linear radio structures showed strong evidence for interactions between the radio 
structures and the emission line gas in the NLR occurring on scales of $\la$1 kpc (e.g. \citealt{Axon98}). In QSO2s, the nuclear 
outflow in Mrk\,477, which is thought to be triggered by the triple radio source present in this galaxy, has an estimated size of 
several hundreds of parsecs \citep{Heckman97,Villar15}, and recent observations of different samples of luminous QSO2s at 
z$<$0.6, some of them with relatively high radio luminosities, reveal compact ionized outflow sizes of $<$1--2 kpc
\citep{Villar16,Karouzos16,Husemann16}. The case of powerful radio galaxies and quasars is different, with 
radio-induced outflows that can extend up to several kpc, even outside the galaxy boundaries \citep{Tadhunter94,Villar99}.

In the case of the H$_2$ lines we do not find evidence for broad components in the nuclear spectrum, and the velocity map 
shows a dominant rotation pattern. The only possible signatures of a molecular outflow in the Teacup are the
nuclear narrow components blueshifted by -50 km~s$^{-1}$ on average relative to the systemic velocity. In any case, the bulk of the 
H$_2$ emission comes from the rotating gas distribution shown in Figure \ref{fig5}, and only a small percentage would be outflowing.  
Deeper observations and a more accurate determination of the systemic velocity of the galaxy are needed to confirm the presence of a nuclear
molecular outflow. 

The case of the Teacup is very different from that of its low-luminosity counterpart IC\,5063. This Seyfert 2 galaxy has a similar
redshift and radio-power as the Teacup but its NIR spectrum shows wings in the H$_2$ profiles as broad as those of the Br$\gamma$ line. 
Furthermore, the H$_2$ emission shows a peak in brightness co-spatial with the radio lobe, which \citet{Tadhunter14} interpreted as 
being due to gas cooling and forming molecules behind a jet-induced shock. The results presented here suggest a different scenario
in the case of the Teacup, demonstrating that we do not always see broad molecular lines in the case of galaxies with 
strong jet-cloud interactions and ionized outflows.

This non-detection 
of broad H$_2$ components in the Teacup could be explained by a
two-stage quasar wind scenario \citep{Lapi05,Menci08,Zubovas12,Faucher12}. Fast winds accelerated by the AGN interact with 
the galaxy ISM, reaching temperatures $\ge10^7$ K. Once the gas cools down to $\sim$10$^4$ K, it starts to emit warm ionized 
and coronal lines such as [O III], Pa$\alpha$, and [Si VI], but further cooling is necessary for the gas to emit in H$_2$.

\subsection{Emission-line structures}

The bulk of the Pa$\alpha$ emission is dominated by the nuclear component, and beyond,
the NE bubble and SW fan expand in opposite directions with projected velocities $\pm$300 km~s$^{-1}$ (see Figure \ref{fig8}). 
This is entirely consistent with the [O III] velocity map reported by \citealt{Harrison15} in a much larger FOV
($\sim$25\arcsec$\times$25\arcsec). 

We extracted spectra at different locations within the emission line structures and we found 
double-peaked Pa$\alpha$ profiles in the SW fan with FWHMs=100--500 km~s$^{-1}$ which are indicative of disturbed kinematics. 
In contrast, in the NE bubble only one Gaussian is needed to reproduce the Pa$\alpha$ lines, but we measure FWHMs=300 km~s$^{-1}$ 
at two different positions. 
For comparison, the extended
non-outflowing ionized gas detected in the most dynamically disturbed mergers with nuclear activity show FWHM$<$250 km~s$^{-1}$ 
\citep{Bellocchi13}.
According to this comparison with merger dynamics, the detection of turbulent gas in the Teacup is confirmed in apertures N1, N2, S1, and S2, 
located at distances between 2.8 and 5.6 kpc from the AGN, which correspond to $\sim$6.5 and $\sim$13.5 times half the seeing size. 


Furthermore, in both the bubble and the fan we find tentative detection of 
a a very broad Pa$\alpha$ component of FWHM$\sim$3000 km~s$^{-1}$. 
These broad components are blueshifted relative to systemic, with Vs ranging between -60 and -310 km~s$^{-1}$.
The velocity shifts that we measure for the very broad components and also for the narrow components (see Table \ref{tab4}) 
are consistent with those recently reported by \citet{Keel17} using GMOS integral field spectroscopy. They claim that asymmetric 
[O III] profiles are present in different locations within the NE bubble, reaching maximum velocities of $\pm$1000 km~s$^{-1}$. 
Unfortunately, \citet{Keel17} did not characterize the lines profiles quantitatively, so we do not know if they are detecting the 
very broad components that we see in our SINFONI data. Our interpretation is 
that, if these broad components are real, we would be detecting highly turbulent gas in a huge outflow that has cooled sufficiently
to emit hydrogen recombination lines and be observable in the NIR. 
Nonetheless, independently of whether or not these broad 
components are real, the large scale gas (up to 5.6 kpc from the nucleus) shows turbulent kinematics that can be explained by 
the action of the outflow.


Finally, the H$_2$ maps presented here reveal a rotation pattern whose axis is misaligned with that of the narrow Pa$\alpha$
emission,
suggesting that the molecular gas is not coupled with the velocity distribution shown by the ionized gas. 
The rotating distribution of molecular gas detected in the Teacup 
could be the quasar-luminosity counterpart of the 100 pc-scale circumnuclear disks (CNDs) observed in nearby Seyfert galaxies using
both NIR \citep{Hicks13} and sub-mm observations \citep{Garcia14}. These CNDs rotate with similar velocities to those measured here 
$\sim$200--300 km~s$^{-1}$ (e.g. \citealt{Helfer03,Garcia05}). Furthermore they are not present in matched samples of quiescent
galaxies \citep{Hicks13}, indicating that they constitute a key element in the feeding of active SMBHs. 


\section{Conclusions}

We have characterized the K-band emission-line spectrum of the radio-quiet QSO2 known 
as the Teacup galaxy. Thanks to the IFU capabilities of SINFONI we have not only studied the nuclear region of the galaxy, but also
the Pa$\alpha$ emission of its kpc-scale emission-line structures (i.e. the NE bubble and the SW fan) within a $\sim$9\arcsec$\times$9\arcsec~FOV. 
Our major conclusions are as follows:

\begin{itemize}

\item The nuclear K-band spectrum of the Teacup reveals the presence of a blueshifted component of 
FWHM$\sim$1600--1800 km~s$^{-1}$ in the hydrogen recombination lines and also in the coronal line [Si VI]$\lambda$1.963. 
Therefore, we confirm the presence of the nuclear ionized outflow previously 
detected from optical spectra, and we reveal its coronal counterpart.

\item The FWHM of the NIR lines associated with the nuclear ionized outflow are larger than those of their optical counterparts. 
This is consistent with the idea that, because of the lower extinction in the NIR, we can trace the outflow closer to its origin. 

\item Both the ionized and coronal nuclear outflows are spatially resolved, with seeing-deconvolved radial sizes of 
1.1$\pm$0.1 and 0.9$\pm$0.1 kpc along the radio axis (PA=72\degr--74\degr). This suggests that the radio jet
could have triggered the nuclear outflow. 

\item We find kinematically disrupted ionized gas (FWHM$>$250 km~s$^{-1}$) 
at up to 5.6 kpc from the AGN, which can be naturally explained by the action of the outflow.

\item The narrow component of [Si VI] is redshifted by V$_s$=54$\pm$11 km~s$^{-1}$ with respect to the 
systemic velocity, 
which is not the case for any other emission line in the K-band spectrum of the Teacup. This 
indicates that the coronal region is not co-spatial with the NLR.


\item In the case of the H$_2$ lines, we do not require a broad component to reproduce the profiles seen in the nuclear spectrum, 
but the narrow lines 
are blueshifted by $\sim$50 km~s$^{-1}$ on average from the galaxy systemic velocity. This could be an indication of the 
presence of a molecular outflow, although additional observations are required to confirm this. 

\item The H$_2$ maps reveal a rotating structure 
oriented roughly perpendicular to the radio jet and the broad Pa$\alpha$ and [Si VI] maps. This molecular gas structure could be 
the quasar-luminosity 
equivalent of the 100 pc-scale CNDs detected in Seyfert galaxies. 

\item We report tentative detection of very broad Pa$\alpha$ components (FWHM$\sim$3000 km~s$^{-1}$) at 
different locations across the NE bubble and SW fan (at up to 5.6 kpc from the AGN). If confirmed, such extremely turbulent components could be hot shocked 
gas that has cooled sufficiently to 
be observable in Pa$\alpha$. 

\end{itemize}

\section*{Acknowledgments}

Based on observations made with ESO Telescopes at the Paranal Observatory under programme ID 094.B-0189(A).
First, we would like to acknowledge the constructive feedback and suggestions of the referee.
CRA acknowledges the Ram\'on y Cajal Program of the Spanish Ministry of Economy and Competitiveness through project 
RYC-2014-15779 and the Spanish Plan Nacional de Astronom\' ia y Astrofis\' ica under grant AYA2016-76682-C3-2-P.
JPL acknowledges support from the Science and Technology Facilities Council (STFC) grant ST/N002717/1 and 
the Spanish Plan Nacional de Astronom\' ia y Astrofis\' ica under grant AYA2012-39408-C02-01.
MVM acknowledges support from the Spanish Ministerio de Econom\' ia y Competitividad through the grants AYA2012-32295 and AYA2015-64346-C2-2-P. 
PSB acknowledges support from FONDECYT through grant 3160374.
The authors acknowledge the data analysis facilities provided by the Starlink Project, 
which is run by CCLRC on behalf of PPARC.
We finally acknowledge Jos\'e Antonio Acosta Pulido, Santiago Garc\' ia Burillo, Richard Davies, 
Erin Hicks, and Clive Tadhunter for useful comments that have substantially contributed to improve this work.



\label{lastpage}


\begin{thebibliography}{}

\bibitem[Appenzeller \& Wagner(1991)]{Appenzeller91} Appenzeller, I., Wagner, S. J. 1991, A\&A, 250, 57
\bibitem[Axon et al.(1998)]{Axon98} Axon, D. J., Marconi, A., Capetti, A., Macchetto, F. D., Schreier, E., Robinson, A. 1998, ApJ, 496,
L75
\bibitem[Bellocchi et al.(2013)]{Bellocchi13} Bellocchi, E., Arribas, S., Colina, L., Miralles-Caballero, D. 2013, A\&A, 557, 59
Cabrera-Lavers, A. 2017, MNRAS, 466, 3887
\bibitem[Black \& van Dishoeck(1987)]{Black87} Black J. H., van Dishoeck E. F. 1987, ApJ, 322, 412
\bibitem[Brotherton et al.(1994)]{Brotherton94} Brotherton, M. S., Wills, B. J., Francis, P. J., Steidel, C. C. 1994, ApJ, 430, 495
\bibitem[Burtscher et al.(2015)]{Burtscher15} Burtscher, L., et al. 2015, A\&A, 578, 47
\bibitem[Croton et al.(2006)]{Croton06} Croton, D. J., et al. 2006, MNRAS, 365, 11
\bibitem[Dale et al.(2005)]{Dale05} Dale, D. A., Sheth, K., Helou, G., Regan, M. W., H\"uttemeister, S. 2005, AJ, 129, 2197
\bibitem[Davies et al.(2014)]{Davies14} Davies, R. I., et al. 2014, ApJ, 792, 101
\bibitem[Denney(2012)]{Denney12} Denney, K. D. 2012, ApJ, 759, 44
\bibitem[Di Matteo et al.(2005)]{diMatteo05} Di Matteo, T., Springel, V., Hernquist, L. 2005, Nature, 433, 604
\bibitem[Draine(1989)]{Draine89} Draine, B. T. 1989, 22nd ESLAB Symp., Infrared Spectroscopy in Astronomy, ed. B. H. Kaldeich 
(ESA SP-290; Paris: ESA), 93
\bibitem[Emonts et al.(2016)]{Emonts16} Emonts, B. H. C., Morganti, R., Villar-Mart\' in, M., Hodgson, J., Brogt, E., Tadhunter, C. N., 
Mahony, E., Oosterloo, T. A. 2016, A\&A, 596, 19
\bibitem[Emonts et al.(2014)]{Emonts14} Emonts, B. H. C., Piqueras-L\'opez, J., Colina, L., Arribas, S., Villar-Mart\' in, M., 
Pereira-Santaella, M., Garcia-Burillo, S., Alonso-Herrero, A. 2014, A\&A, 572, 40
\bibitem[Fabian(2012)]{Fabian12} Fabian, A. C. 2012, ARA\&A, 50, 455
\bibitem[Faucher-Gigu\'ere \& Quataert(2012)]{Faucher12} Faucher-Gigu\'ere, C. A., Quataert, E. 2012, MNRAS, 425, 605
\bibitem[Fiore et al.(2017)]{Fiore17} Fiore, F., et al. 2017, A\&A, in press, arXiv:1702.04507
\bibitem[Garc\' ia-Burillo et al.(2005)]{Garcia05} Garc\' ia-Burillo, S., Combes, F., Schinnerer, E., Boone, F., Hunt, L. K. 2005,
A\&A, 441, 1011
\bibitem[Gagne et al.(2014)]{Gagne14} Gagne, J. P., Crenshaw, D. M., Kraemer, S. B., et al. 2014, ApJ, 792, 72
\bibitem[Garc\' ia-Burillo et al.(2014)]{Garcia14} Garc\' ia-Burillo, S., et al. 2014, A\&A, 567, 125
\bibitem[Gelbord et al.(2009)]{Gelbord09} Gelbord, J. M., Mullaney, J. R., Ward, M. J. 2009, MNRAS, 397, 172
\bibitem[(H2015)]{Harrison15} Harrison, C. M., Thomson, A. P., Alexander, D. M., Bauer, F. E., Edge, A. C., 
Hogan, M. T., Mullaney, J. R., Swinbank, A. M. 2015, ApJ, 800, 45
\bibitem[Harrison et al.(2014)]{Harrison14} Harrison, C. M., Alexander, D. M., Mullaney, J. R., Swinbank, A. M. 
2014, MNRAS, 441, 3306
\bibitem[Heckman et al.(1997)]{Heckman97} Heckman, T. M., Gonz\'alez-Delgado, R., Leitherer, C., Meurer, G. R., Krolik, J., Wilson, A. S., Koratkar, A., Kinney, A.
1997, ApJ, 482, 114
\bibitem[Helfer et al.(2003)]{Helfer03} Helfer, T. T., Thornley, M. D., Regan, M. W., Wong, T., Sheth, K., Vogel, S. N., 
Blitz, L., Bock, D. C.-J. 2003, ApJS, 145, 259
\bibitem[Hicks et al.(2013)]{Hicks13} Hicks, E. K. S., Davies, R. I., Maciejewski, W., Emsellem, E., Malkan, M. A., 
Dumas, G., M\"uller-S\'anchez, F., Rivers, A. 2013, ApJ, 768, 107
\bibitem[Hollenbach \& McKee(1989)]{Hollenbach89} Hollenbach D., McKee C. F. 1989, ApJ, 342, 306 
\bibitem[Hummer \& Storey(1987)]{Hummer87} Hummer, D. G., Storey, P. J. 1987, MNRAS, 224, 801
\bibitem[Humphrey et al.(2010)]{Humphrey10} Humphrey, A., et al. 2010, MNRAS, 408, L1
\bibitem[Husemann et al.(2016)]{Husemann16} Husemann, B., Scharw\"achter, J., Bennert, V. N., Mainieri, V., Woo, J.-H., Kakkad, D.
2016, A\&A, 594, A44
\bibitem[Karouzos et al.(2016)]{Karouzos16} Karouzos, M., Woo, J.-H., Bae, H.-J. 2016, ApJ, 819, 148
\bibitem[Keel et al.(2017)]{Keel17} Keel, W. C., et al. 2016, ApJ, 835, 256
\bibitem[Keel et al.(2015)]{Keel15} Keel, W. C., et al. 2015, AJ, 149, 155
\bibitem[Keel et al.(2012)]{Keel12} Keel, W. C., et al. 2012, MNRAS, 420, 878
\bibitem[Korista \& Ferland(1989)]{Korista89} Korista, K. T., Ferland, G. J. 1989, ApJ, 343, 678
\bibitem[Lal \& Ho(2010)]{Lal10} Lal, D. V., Ho, L. C. 2010, AJ, 139, 1089
\bibitem[Landt et al.(2015)]{Landt15} Landt, H., Ward, M. J., Steenbrugge, K. C., Ferland, G. J. 2015, MNRAS, 449, 3795
\bibitem[Lapi et al.(2005)]{Lapi05} Lapi, A., Cavaliere, A., Menci, N. 2005, ApJ, 619, 60
\bibitem[Liu et al.(2013)]{Liu13} Liu, G., Zakamska, N., Greene, J. E., Nesvadba, N. P. H., Liu, X. 2013, MNRAS, 430, 2327
\bibitem[Maloney et al.(1996)]{Maloney96} Maloney P. R., Hollenbach D. J., Tielens A. G. G. M. 1996, ApJ, 466, 561
\bibitem[Mazzalay et al.(2013)]{Mazzalay13} Mazzalay X., et al. 2013, MNRAS, 428, 2389
\bibitem[Menci et al.(2008)]{Menci08} Menci, N., Fiore, F., Puccetti, S., Cavaliere, A. 2008, ApJ, 686, 219
\bibitem[Mouri(1994)]{Mouri94} Mouri, H. 1994, ApJ, 427, 777
\bibitem[Mullaney et al.(2013)]{Mullaney13} Mullaney J. R., Alexander, D. M., Fine, S., Goulding, A. D., Harrison, C. M., 
Hickox, R. C. 2013, MNRAS, 433, 622
\bibitem[Mullaney et al.(2009)]{Mullaney09} Mullaney J. R., Ward M. J., Done, C., Ferland, G. J., Schurch, N. 2009, MNRAS, 394, L16
\bibitem[Mullaney \& Ward(2008)]{Mullaney08} Mullaney J. R., Ward M. J., 2008, MNRAS, 385, 53
\bibitem[M\"uller-S\'anchez et al.(2016)]{Muller16} M\"uller-S\'anchez, F., Comerford, J., Stern, D., Harrison, F. A. 
2016, ApJ, 830, 50
\bibitem[M\"uller-S\'anchez et al.(2011)]{Muller11} M\"uller-S\'anchez, F., Prieto, M. A., Hicks, E. K. S., Vives-Arias, H., 
Davies, R. I., Malkan, M., Tacconi, L. J., Genzel, R. 2011, ApJ, 739, 69
\bibitem[M\"uller-S\'anchez et al.(2006)]{Muller06} M\"uller-S\'anchez, F., Davies, R. I., Eisenhauer, F., Tacconi, L. J., 
Genzel, R., Sternberg, A. 2006, A\&A, 454, 481
\bibitem[Penston et al.(1984)]{Penston84} Penston M. V., Fosbury R. A. E., Boksenberg A., Ward M. J., Wilson A. S.,
1984, MNRAS, 208, 347
\bibitem[Piqueras L\'opez et al.(2016)]{Piqueras16}  Piqueras L\'opez, J., Colina, L., Arribas, S., Pereira-Santaella, M. 
Alonso-Herrero, A., 2016, A\&A, 590, A67
\bibitem[Piqueras L\'opez et al.(2012)]{Piqueras12}  Piqueras L\'opez, J., Colina, L., Arribas, S., Alonso-Herrero, A., 
Bedregal, A. G. 2012, A\&A, 546, A64
\bibitem[Ramos Almeida et al.(2009)]{Ramos09} Ramos Almeida, C., P\'erez Garc\' ia, A. M., Acosta-Pulido, J. A. 2009, ApJ, 694,
1379
\bibitem[Reunanen et al.(2002)]{Reunanen02} Reunanen J., Kotilainen J. K., Prieto M. A. 2002, MNRAS, 331, 154
\bibitem[Reyes et al.(2008)]{Reyes08} Reyes, R., et al. 2008, AJ, 136, 2373 
\bibitem[Riffel et al.(2006)]{Riffel06} Riffel, R., Rodr\' iguez-Ardila, A., Pastoriza, M. G. 2006, A\&A, 457, 61
\bibitem[Rodr\' iguez-Ardila et al.(2011)]{Rodriguez11} Rodr\' iguez-Ardila, A., Prieto, M. A., Portilla, J. G., Tejeiro, J. M.
2011, ApJ, 743, 100
\bibitem[Rodr\' iguez-Ardila et al.(2006)]{Rodriguez06b} Rodr\' iguez-Ardila, A., Prieto, M. A., Viegas, S., Gruenwald, R. 2006, 
ApJ, 653, 1098
\bibitem[Rodr\' iguez-Ardila \& Mazzalay(2006)]{Rodriguez06} Rodr\' iguez-Ardila, A., Mazzalay, X. 2006, MNRAS, 367, L57
\bibitem[Rose et al.(2011)]{Rose11} Rose, M., Tadhunter, C. N., Holt, J., Ramos Almeida, C., Littlefair, S. 2011, MNRAS, 414, 3360
\bibitem[Tadhunter et al.(2014)]{Tadhunter14} Tadhunter, C., Morganti, R., Rose, M., Oonk, J. B. R., Oosterloo, T. 2014, 
Nature, 511, 440
\bibitem[Tadhunter et al.(1994)]{Tadhunter94} Tadhunter, C., Shaw, M., Clark, N., Morganti, R. 1994, A\&A, 288, L21

\bibitem[Villar-Mart\' in et al.(2016)]{Villar16} Villar-Mart\' in, Arribas, S., Emonts, B., Humphrey, A., Tadhunter, C., 
Bessiere, P., Cabrera Lavers, A., Ramos Almeida, C. 2016, MNRAS, 460, 130
\bibitem[Villar Mart\' in et al.(2015)]{Villar15} Villar Mart\' in, M., Bellocchi, E., Stern, J., Ramos Almeida, C., Tadhunter, C., Gonz\'alez Delgado, R. 2015,
MNRAS, 454, 439
\bibitem[Villar-Mart\' in et al.(2013)]{Villar13} Villar-Mart\' in, M., et al. 2013, MNRAS, 434, 978
\bibitem[Villar-Mart\' in et al.(2014)]{Villar14} Villar-Mart\' in, M., Emonts, B., Humphrey, A., Cabrera Lavers, A., Binette, L.
2014, MNRAS, 440, 3202
\bibitem[Villar-Mart\' in et al.(2011)]{Villar11} Villar-Mart\' in, M., Humphrey, A., Gonz\'alez Delgado, R., Colina, L., Arribas, S. 
2011, MNRAS, 418, 2032
\bibitem[Villar-Mart\' in et al.(1999)]{Villar99} Villar-Mart\' in, M., Tadhunter, C., Morganti, R., Axon, D., Koekemoer, A. 1999, MNRAS, 307, 24
\bibitem[Zakamska \& Greene(2014)]{Zakamska14} Zakamska, N. L., Greene, J. E. 2014, MNRAS, 442, 784
\bibitem[Zubovas \& King(2012)]{Zubovas12} Zubovas, K., King, A. 2012, ApJ, 745, L34


\end{thebibliography}
\end{document}